\documentclass[journal]{IEEEtran}

\usepackage{layout}
\usepackage[letterpaper, left=.92in, right=.92in, bottom=.7in, top=0.70in]{geometry}

\def\bb0{{\mathbb{0}}}

\def\bb{{\boldsymbol{b}}}

\def\bh{{\boldsymbol{h}}}

\def\bu{{\boldsymbol{u}}}
\def\bv{{\boldsymbol{v}}}

\def\bx{{\boldsymbol{x}}}
\def\by{{\boldsymbol{y}}}
\def\bz{{\boldsymbol{z}}}
\def\b0{{\boldsymbol{0}}}

\def\bD{{\boldsymbol{D}}}

\def\bF{{\boldsymbol{F}}}

\def\bH{{\boldsymbol{H}}}
\def\bI{{\boldsymbol{I}}}

\def\bR{{\boldsymbol{R}}}

\def\s{{\mathrm{s}}}

\def\b{{\mathrm{b}}}

\def\r0{{\mathbf{0}}}

\def\bbP{{\mathbb{P}}}

\def\cH{\mathcal{H}}
\def\cI{\mathcal{I}}

\def\cN{\mathcal{N}}
\def\cO{\mathcal{O}}

\def\cY{\mathcal{Y}}

\def\bSigma{\bm \Sigma}

\def\sfy{{\mathsf{y}}}

\def\bsfx{{\bm{\mathsf{x}}}}
\def\bsfy{{\bm{\mathsf{y}}}}

\def\bsf0{{\bm{\mathsf{0}}}}

\def\c{{\rm c}} % speed of light

\def\Nt{{N_{\mathrm{t}}}}
\def\Nr{{N_{\mathrm{r}}}}

\def\Nmax{{N_{\mathsf{max}}}}
\def\Nmin{{N_{\mathsf{min}}}}

\def\Pt{{P_{\mathrm{t}}}}

\def\N0{{N_{\mathrm{0}}}}

\def\j{\mathrm{j}}

\newcommand{\Ncgauss}{\mathcal{N}_{\mathbb{C}}}

\def\rmd{\mathrm{d}}

\def\thetat{\theta_{\rm t}}
\def\thetar{\theta_{\rm r}}

\newcommand{\dt}{{d_{\rm t}}}
\newcommand{\dr}{{d_{\rm r}}}

\def\bsf{{\boldsymbol{s}_\mathrm{f}}}

\newcommand{\be}{\begin{equation}}
\newcommand{\ee}{\end{equation}}
\newcommand{\bal}{\begin{align}}
\newcommand{\eal}{\end{align}}
\def\tr {{\rm tr}}

\def\C{{\mathsf{C}}}

\def\SNR    {{\mathsf{SNR}}}

\def\ebno   {\frac{E_{\rm b}}{N_0}}
\def \ebnoinline {\frac{E_{\rm b}}{N_0}}
\def\ebnomin  {\ebno_{\rm \min}}
\def\ebnomininline   {\ebnoinline_{\rm min}}

\def\dB {{|_{\mathrm{\scriptscriptstyle dB}}}}

\usepackage{graphicx}
\usepackage{amsmath}
\usepackage{subfig}
\usepackage{mathtools}
\usepackage{epsfig}
\usepackage{float}
\usepackage{lipsum}
\usepackage{stfloats}
\usepackage{array}
\usepackage{amssymb}
\usepackage{cite}
\usepackage{url}
\usepackage{amsthm}
\usepackage{algorithm}
\usepackage{algorithmic}
\usepackage{multirow}
\usepackage{upgreek}
\usepackage{diagbox}
\usepackage{dsfont}
\usepackage{xifthen}% provides \isempty test

\normalsize
\allowdisplaybreaks

\newcommand{\GD}[1][]{3
	\ifthenelse{\isempty{#1}}%
	{\text{G}_{i,j}^k}% if #1 is empty
	{{(\text{G}_{i,j}^k)}^{#1}}% if #1 is not empty
}

\theoremstyle{definition}

\usepackage{layout}
\usepackage{bm}
\usepackage{comment}
\usepackage{color,soul}
\usepackage{makecell}
\usepackage{amssymb}

\def\*#1{\mathbf{#1}}
\def\w*#1{\widehat{#1}}

\def\s*#1{\mathsf{#1}}
\def\S*#1{\bm{\mathsf{#1}}}

\def\c*#1{\mathcal{#1}}
\def\C*#1{\bm{\mathcal{#1}}}
\def\T*#1{\text{#1}}
\newcommand{\E}{\mathbb{E}}

\begin{document}

%\title{Switched Beamforming and Equiprobable Signaling for 1-Bit MIMO}
\title{1-Bit MIMO for Terahertz Channels
\thanks{A. Lozano is with Univ. Pompeu Fabra, 08018 Barcelona (e-mail: angel.lozano@upf.edu). His work is supported by the European Research Council under the H2020 Framework Programme/ERC grant agreement 694974, by MINECO's Projects RTI2018-102112 and RTI2018-101040, and by ICREA.
Parts of this paper were presented at the 2021 Int'l ITG Workshop on Smart Antennas \cite{WSA2021}.}
}
\author{\IEEEauthorblockN{Angel~Lozano}, {\it Fellow,~IEEE}}
\maketitle

\maketitle

\begin{abstract}
This paper tackles the problem of single-user multiple-input multiple-output communication with \mbox{1-bit} digital-to-analog and analog-to-digital converters. 
With the information-theoretic capacity as benchmark, the complementary strategies of beamforming and equiprobable signaling
 are contrasted in the regimes of operational interest, and the ensuing spectral efficiencies are characterized.
Various canonical channel types are considered, with emphasis on line-of-sight settings under both spherical and planar wavefronts, respectively representative of short and long transmission ranges at mmWave and terahertz frequencies. In all cases, a judicious combination of beamforming and equiprobable signaling is shown to operate within a modest gap from capacity.
%corresponding to extreme settings that encompass the range of possibilities expected for terahertz communication.
\end{abstract}

%REFERENCES ON THz CHANNEL MODELING \cite{6574880,han2014multi}

\section{Introduction}

As they evolve, wireless systems seek to provide ever faster bit rates and lower latencies, and a key enabler for these advances in the increase in bandwidth. From 1G to 5G, the spectrum devoted to wireless communication has surged from a handful of MHz to multiple GHz, roughly three orders of magnitude, and this growth is bound to continue as new mmWave bands open up and inroads are made into the terahertz realm \cite{4455844,akyildiz2014terahertz,elayan2019terahertz,8732419,9216613}.

Besides bandwidth, another key resource is power. Leaving aside the power spent in duties unrelated to communication, the power consumed by a device
%that transmits and receives
can be partitioned as $\Pt/\eta + P_{\sf ADC} + P_{\sf other}$ where $\Pt$ is the power radiated by the transmitter, $\eta$ is the efficiency of the corresponding power amplifiers, $P_{\sf ADC}$ is the power required by the receiver's analog-to-digital (ADC) converters, and $P_{\sf other}$ subsumes everything else (including oscillators, filters, the transmitter's digital-to-analog (DAC) converters, and the receiver's low-noise amplifier). % such that
%\be
%\frac{\Pt}{\eta} + P_{\sf ADC} + P_{\sf other} \leq P
%\ee
%where $P$ is the power budgeted for communication purposes.
With $B$ denoting the bandwidth and $b$ the resolution in bits, each ADC satisfies
\be
P_{\sf ADC} = \mathsf{FoM} \, B \, \kappa^b
\label{LMessi}
\ee
where $\mathsf{FoM}$ is a figure of merit and $\kappa$ ranges between two and four \cite{lee2008analog,murmann2016adc}.
%FOR HIGH RESOLUTION IT IS $P_{\sf ADC} = \epsilon B 4^b$.
%WE HAVE IDENTIFIED SAMPLING RATE WITH BANDWIDTH

Power consumption has traditionally been dominated by $\Pt / \eta$ and thus high resolutions ($b=8$--$12$ bits) could be employed. In 1G and 2G, a higher $\eta$ was facilitated by the adoption of (respectively analog and digital) signaling formats tolerant of nonlinear amplification, but after 2G this took a backseat to spectral efficiency. Linearity has since reigned, despite the lower $\eta$, as $\Pt / \eta$ was well within the power budget of devices for the desirable $\Pt$.

%THE TENSION BETWEEN BANDWIDTH AND POWER BEING THE LIMITING FACTOR HAS OSCILLATED OVER TIME, WITH IMPACT ON THE DESIGN OF SYSTEMS. IN 1G-2G, POWER WAS DEEMED MORE LIMITING AND NONLINEAR SIGNALING CHOSEN. WITH THE EXPLOSION IN COST OF SPECTRUM, BY 3G BANDWIDTH WAS PERCEIVED AS THE LIMITING FACTOR. LIKEWISE IN 4G, DESPITE THE ALLOWANCE OF SF-FDMA FOR THE UPLINK. IN 5G, WITH HUGE CHUNKS OF mmWAVE SPECTRUM OPENING UP, THE EMPHASIS IS SHIFTING BACK TO POWER EFFICIENCY. THIS SHOULD INTENSEIFY MOVING FORWARD.

The advent of 5G, with the move up to mmWave frequencies and the enormous bandwidths therein, is a turning point in the sense of $P_{\sf ADC}$ ceasing to be secondary, and this can only accelerate moving forward \cite{skrimponis2020power}.
Consider this progression: with $b=10$ at a typical 4G bandwidth of $B=20$ MHz, $P_{\sf ADC}$ is only a few milliwatts; for $B=2$ GHz, it is already on the order of a watt; and for $B=20$ GHz, it would reach roughly 10 watts. % all while $\Pt \leq 200$ mW.)
Indeed, as $B$ continues to grow, $P_{\sf ADC}$ is bound to swallow up the entire power budget of portable devices unless $\mathsf{FoM}$ or $b$ change.
But $\mathsf{FoM}$ is approaching a fundamental limit \cite{murmann2016adc}. Moreover, while holding steady up to about $B=100$ MHz, $\mathsf{FoM}$ drops sustainedly after that mark, which is coincidentally the largest 4G bandwidth. Inevitably then, $b$ has to decrease and, ultimately, it should reach $b=1$, to
drastically curb the power consumption and to further enable dispensing with automatic gain control at the receiver while simplifying the data pipeline between the ADCs and the baseband processing \cite{o2005ultra}.

While 1-bit ADCs curb the spectral efficiency at 1 bit per dimension, the vast bandwidths thereby rendered possible make it exceedingly beneficial. Going from $b=10$ down to $b=1$ cuts the spectral efficiency by a factor of 2--3, but in exchange $B$ can grow by as much as 1000 under the same $P_{\sf ADC}$; the net benefits in bit rate and latency are stupendous. Spectral efficiency is then best recovered by expanding the number of antennas, which $P_{\sf ADC}$ is only linear in. % rather than exponentially.
This naturally leads to multiple-input multiple-output (MIMO) arrangements with 1-bit ADCs.

%FINAL ARGUMENT TO ADD: ADDITIONAL ANTENNAS ARE A MUCH BETTER WAY TO RECOVER SPECTRAL EFFICIENCY THAN ADDITIONAL RESOLUTION BITS. SINCE ADDITIONAL ANTENNA  INCREASE POWER CONSUMPTION ROUGHLY LINEARLY, THEY ARE EQUIVALENT TO MORE BANDWIDTH. FORMALIZE AND EXEMPLIFY THIS, AS LEAD-IN TO MIMO.

Although 1-bit ADCs at the receiver do not necessarily entail 1-bit DACs at the transmitter, and in some cases the spectral efficiency could improve somewhat with richer DACs, it is inviting to take the opportunity and adopt 1-bit transmit signals. This not only minimizes the DAC power consumption---somewhat lower than its ADC's counterpart, yet also considerable \cite{nasri2017700,olieman2015interleaved,shu2018calibration}---but it enables the power amplifiers to operate in nonlinear regimes where $\eta$ is higher.
%Then, $\Pt/\eta + P_{\sf ADC}$ becomes optimized and the emphasis can shift to improving $P_{\sf other}$.  
%The DACs, while somewhat less power hungry than the ADCs, also contribute considerably \cite{nasri2017700,olieman2015interleaved,shu2018calibration}.
    
%LOOSE ANALOGY: RISC COMPUTING, WHICH IS WHAT DOMINATES ALL PORTABLE DEVICES: SIMPLER OPERATIONS, MANY MORE OF WHICH CAN BE CONDUCTED PER UNIT TIME, ENDS OF BEING BETTER.

Altogether, 1-bit MIMO architectures might feature prominently in future wireless systems, and not only for mmWave or terahertz operation: these architectures are also a sensible way forward for lower-frequency extreme massive MIMO, with antenna counts in the hundreds or even thousands \cite{de2020non}.
All this interest is evidenced by the extensive literature on transmission strategies and the ensuing performance with 1-bit converters at the transmitter or receiver only (see \cite{nossek2006capacity,mezghani2007modified,mezghani2008analysis,singh2009limits,singh2009multi,7458830,7307134,wang2014multiuser,mo2015capacity,7600443,li2017channel,jacobsson2017throughput,rassouli2018gaussian,8437510,mezghani2020low,jacobsson2016nonlinear,mezghani2007ultra,mezghani2012capacity,8754755,8487043,8811616,8331077,8103022,7472304,8462805,8010806,zeitler2012low,1683157,6545291,mo2017hybrid,liang2016mixed} and references therein),
and by the smaller but growing body of work that considers 1-bit converters at both ends \cite{gao2017power,gao2018beamforming,mezghani2009transmit,usman2016mmse,kakkavas2016weighted,7569655,guerreiro2016use,li2017downlink,gao2018capacity,7967843,7946265,nam2019capacity,Eusipco21,bazrafkan2020asymptotic}. 
%A host of issues that are thoroughly understood for full-resolution settings must tackled anew. % for the 1-bit realm in these contributions.
Chief among the difficulties in this most stringent case stand (\emph{i}) computing the information-theoretic performance limits for moderate and large antenna counts, and (\emph{ii}) precoding to generate signals that can approach those limits. % in different types of channels.
%Both these aspects, thoroughly understood for full-resolution settings, must be tackled anew and become

On these fronts, and concentrating on single-user MIMO,
%and building on some of the prior works \cite{mezghani2007ultra,mezghani2012capacity,gao2017power,gao2018beamforming},
this paper has a two-fold objective:
\begin{itemize}
\item To provide analytical characterizations of the performance of beamforming and equiprobable signaling, two transmission strategies that are information-theoretically motivated and complementary. %in the regimes of operational interest.
\item To show that a judicious combination of these strategies suffices to operate within a modest gap from the 1-bit capacity in various classes of channels
of high relevance, foregoing general precoding solutions.
\end{itemize}

%Section \ref{calor1} sets out the signal and channel models, Section~\ref{calor2} then formulates the 1-bit capacity and Section~\ref{calor3} offers a way of putting it on an equal footing with
%its full-resolution counterpart. Sections~\ref{calor4} and \ref{calor5} delve into the performance of beamforming and equiprobable signaling, which Section~\ref{calor6} specializes to various channels of interest before Section~\ref{calor7} concludes the paper.

\section{Signal and Channel Models}
\label{calor1}

\subsection{Signal Model}

Consider a transmitter equipped with $\Nt$ antennas and 1-bit DACs per complex dimension. The receiver, which features $\Nr$ antennas and a 1-bit ADC per complex dimension, observes
\be
\label{Aleix}
\by = \text{sgn} \! \left( \sqrt{\frac{\SNR}{2 \Nt}} \, \bH \bx + \bz \right)
\ee
where the sign function is applied separately to the real and imaginary parts of each entry, such that $y_n \in \{\pm1\pm\j\}$, while $\bH$ is the $\Nr \times \Nt $ channel matrix,
%normalized to have unit-variance entries,
$\bz \sim \Ncgauss({\bf 0}, \bI)$  is the noise, and $\SNR$ is the signal-to-noise ratio per receive antenna. Each entry of the transmit vector $\bx$ also takes the values $\pm1\pm\j$.

Each antenna in the foregoing formulation could actually correspond to a compact subarray, in which case the model subsumes array-of-subarrays structures for the transmitter and/or receiver \cite{Allerton2006,Torkildson:11,Song:152,Lin:16,ISIT} provided $\SNR$ is appropriately scaled. 

For each given $\bH$, the relationship in (\ref{Aleix}) embodies a discrete memoryless channel with $4^{\Nt} \times 4^{\Nr}$ transition probabilities determined by %\cite{gao2017power}
\begin{align}
p_{\by|\bx} & = \prod_{n=0}^{\Nr-1} p_{\Re\{ y_n \} | \bx} \, p_{\Im\{y_n\} | \bx} ,
\end{align}
where the factorization follows from the noise independence per receive antenna and complex dimension.
Each such noise component has variance $1/2$, hence
\begin{align}
p_{\Re\{y_n\} | \bx} ( 1 | \bsfx) & = \text{Pr} \! \left[ \sqrt{\frac{\SNR}{2 \Nt}} \Re\{  \bh_n \bsfx + z_n \} >0   \right] \\
& = \text{Pr} \! \left[ \Re\{ z_n \} > -  \sqrt{\frac{\SNR}{2 \Nt}} \Re\{  \bh_n \bsfx \}   \right] \\
& = Q  \! \left( \! -  \sqrt{\frac{\SNR}{\Nt}} \Re\{\bh_n \bsfx\} \! \right) \label{Magda}
\end{align}
where $\bh_n$ is the $n$th row of $\bH$ (for $n=0,\ldots,\Nr-1$) and $Q(\cdot)$ is the Gaussian Q-function. Similarly,
\be
p_{\Re\{y_n\} | \bx} ( -1 | \bsfx) =  Q  \! \left( \!  \sqrt{\frac{\SNR}{\Nt}} \Re\{\bh_n \bsfx\} \! \right) .
\label{Campins}
\ee
From (\ref{Magda}) and (\ref{Campins}), %we can write
\be
p_{\Re\{y_n\} | \bx} ( \Re\{ \sfy_n \} | \bsfx) = Q  \! \left( \! - \Re\{\sfy_n\} \sqrt{\frac{\SNR}{\Nt}} \Re\{\bh_n \bsfx\} \! \right) 
\ee
and, mirroring it, finally
\begin{align}
p_{\by|\bx}(\bsfy | \bsfx) & = \prod_{n=0}^{\Nr-1} Q  \! \left( \! - \Re\{\sfy_n\} \sqrt{\frac{\SNR}{\Nt}} \Re\{\bh_n \bsfx\} \! \right) \nonumber \\
& \quad \cdot  Q  \! \left( \! - \Im\{\sfy_n\} \sqrt{\frac{\SNR}{\Nt}} \Im\{\bh_n \bsfx\} \! \right) .
\label{Joan}
\end{align}

The transition probabilities correspond to (\ref{Joan}) evaluated for the $4^\Nr$ possible values of $\by$ and the $4^\Nt$ values of $\bx$.
If $\bH$ is known, these transition probabilities can be readily computed. Conversely, if the transition probabilities are known, $\bH$ can be deduced.

The $4^{\Nt}$ transmit vectors $\bx$ can be partitioned into $4^{\Nt - 1}$ quartets,  
each containing four vectors and being invariant under a $90^\circ$ phase rotation of all the entries: from any vector in the quartet, the other three are obtained by repeatedly multiplying by $\j$. Since a $90^\circ$ phase rotation of $\bx$ propagates as a $90^\circ$ phase rotation of $\bH \bx$, and the added noise is circularly symmetric, the four vectors making up each transmit quartet are statistically equivalent and they should thus have the same transmission probability so as to convey the maximum amount of information. %This intuition, formalized for $\Nr=1$ in \cite[lemma 1]{nam2019capacity}, leads to $p_\bx$ being a proper complex distribution, i.e., such that $p_\bx(\bsfx) = p_\bx(\j \bsfx)$ \cite[app. C.1]{Foundations:18}.

Likewise, the set of $4^\Nr$ possible vectors $\by$ can be partitioned into $4^{\Nr - 1}$ quartets, and the four vectors $\by$ within each received quartet are equiprobable.

\subsection{Channel Model}
\label{Carla18}

If the channel is stable over each codeword, then every realization of $\bH$ has operational significance and $\SNR$ is well defined under the normalization $\tr(\bH\bH^*)=\Nt\Nr$. Conversely, if the coding takes place over a sufficiently broad range of channel fluctuations, that significance is acquired in an ergodic sense with $\E \big[ \tr(\bH\bH^*) \big]=\Nt\Nr$ \cite{LozanoJindal2012}.
The following classes of channels are specifically considered.
% extremes in terms of multipath richness.

\paragraph{Line-of-Sight (LOS) with Spherical Wavefronts} 

LOS is the chief propagation mechanism at mmWave and terahertz frequencies, and the spherical nature of the wavefronts is relevant for large arrays and short transmission ranges. For uniform linear arrays (ULAs) \cite{do2020reconfigurable},
\be
\bH = \bD_{\rm rx} \tilde{\bH} \bD_{\rm tx}
\label{Oscarinyu}
\ee
where $\bD_{\rm rx}$ and $\bD_{\rm rx}$ are diagonal matrices with entries
\begin{align}
[\bD_{\rm rx}]_{n,n} & = e^{-j \pi \left[ \frac{2n}{\uplambda} \dr \sin \! \thetar \cos \! \phi + \frac{n^2}{\uplambda D} d^2_{\rm r} \, (1-\sin^2 \! \thetar \cos^2 \! \phi) \right] } \nonumber  \\
[\bD_{\rm tx}]_{m,m} & = e^{-j \pi \left[ \frac{2m}{\uplambda} \dt \sin \! \thetat + \frac{m^2 }{\uplambda D} d^2_{\rm t} \right] }
\label{ventvent}
\end{align}
and with $D$ the range, $\uplambda$ the wavelength, $d_{\rm t}$ and $d_{\rm r}$ the antenna spacings at transmitter and receiver, $\theta_{\rm t}$ and $\theta_{\rm r}$ the transmitter and receiver elevations, and $\phi$ their relative azimuth angle.
%\begin{align}
%    h_{n,m}
%    & = %e^{-j2\pi \frac{D}{\lambda}}  \; 
%    e^{-j \pi \left[ \frac{2n}{\lambda} \dr \sin \! \thetar \cos \! \phi + \frac{n^2}{\lambda D} d^2_{\rm r} \, (1-\sin^2 \! \thetar \cos^2 \! \phi) \right] } \label{eq:nonParallelUlaChannel} \\
%  & \quad \cdot e^{j 2\pi \frac{nm}{\lambda D} \dr \dt \cos \! \thetar \cos \! \thetat} \;  
%  e^{-j \pi \left[ \frac{2m}{\lambda} \dt \sin \! \thetat + \frac{m^2 }{\lambda D} d^2_{\rm t} \right] } \nonumber
%\end{align}
%where $D$ is the range, $\lambda$ is the wavelength, $d_{\rm t}$ and $d_{\rm r}$ are the antenna spacings at transmitter and receiver, $\theta_{\rm t}$ and $\theta_{\rm r}$ are the transmitter and receiver elevation angles, and $\phi$ is their relative azimuth angle.
In turn, $\tilde{\bH}$ is the Vandermonde matrix
\begin{align}
\setlength\arraycolsep{2pt}
\tilde{\bH} = \begin{bmatrix}
    e^{j2\pi\eta \frac{0\times0}{\Nmax}} & \cdots & e^{j2\pi\eta \frac{(N_{\rm t}-1)\times0}{\Nmax}} \\
    \vdots & \ddots & \vdots \\
    e^{j2\pi\eta \frac{0\times (N_{\rm r}-1)}{\Nmax}} & \cdots & e^{j2\pi\eta \frac{(N_{\rm t}-1)\times(N_{\rm r}-1)}{\Nmax}}
    \end{bmatrix}
    \label{MesMessi}
\end{align}
where $\Nmax=\max(\Nt,\Nr)$ while
\be
\eta=\frac{ (\dr \cos \theta_{\rm r}) (\dt  \cos \theta_{\rm t}) \Nmax}{\uplambda D}
\label{eq:nonPa_eta}
\ee
is a parameter that concisely describes any LOS setting with ULAs.

Uniform rectangular arrays can be expressed as the Kronecker product of ULAs, and expressions deriving from (\ref{Oscarinyu}) emerge \cite{Larsson:05}.
For more complex topologies, the entries of $\bH$ continue to be of unit magnitude, but the pattern of phase variations becomes more cumbersome.

\paragraph{LOS with Planar Wavefronts}

For long enough transmission ranges, the planar wavefront counterpart to (\ref{Oscarinyu})
%and (\ref{MesMessi}) are
is obtained by letting $D \to \infty$, whereby the channel becomes rank-1 with
\begin{align}
    h_{n,m} = e^{-\j \frac{2 \pi }{\lambda} (n d_{\rm r} \sin \theta_{\rm r} \cos \phi + m d_{\rm t} \sin \theta_{\rm t} )}.
    \label{antigens}
\end{align}

\paragraph{IID Rayleigh Fading}

In this model, representing situations of rich multipath propagation, the entries of $\bH$ are IID and $h_{n,m} \sim \Ncgauss(0,1)$.

We note that the frequency-flat representation embodied by $\bH$ is congruous for the two LOS channel models, but less so for the IID model, where the scattering would go hand in hand with frequency selectivity over the envisioned bandwidths. The analysis presented for this model intends to set the stage for more refined characterizations that account for the inevitable intersymbol interference.
In fact, even for the LOS channels, over a sufficiently broad bandwidth there is bound to be intersymbol interference because of spatial widening, i.e., because of the distinct propagation delays between the various transmit and receive antennas \cite{8443598,8354789}.

\section{1-Bit Capacity}
\label{calor2}

Denote by $p_1,\ldots,p_{4^{\Nt-1}}$ the activation probabilities of the transmit quartets, such that $\sum_k p_k = 1$
%\be
%\sum_{k=1}^{4^{\Nt-1}} p_k = 1 
%\ee
and $p_\bx(\bsfx_k)=p_k/4$ with $\bsfx_k$ any of the vectors in the $k$th quartet.
Letting $\cH(\cdot)$ indicate entropy, and with all the probabilities conditioned on $\bH$, the mutual information is
\begin{align}
\!\! \cI(\SNR,\bH) %& = I (\bx;\by | \bH) \\
& = \cH(\by) - \cH(\by | \bx) \\
& = \sum_{\ell=1}^{4^{\Nr}} p_{\by}(\bsfy_\ell) \log_2 \frac{1}{p_{\by}(\bsfy_\ell)} - \cH(\by | \bx) \\
& = 4 \! \sum_{\ell=1}^{4^{\Nr-1}} \! p_{\by}(\bsfy_\ell) \log_2 \frac{1}{p_{\by}(\bsfy_\ell)} - \cH(\by | \bx)
\label{NoWiFi2}
\end{align}
where (\ref{NoWiFi2})  follows from the equiprobability of the vectors in each received quartet and $\bsfy_\ell$ is any of the vectors in the $\ell$th such quartet %\footnote{For $\Nr=1$, $p_y=1/4$ and thus $\cH(y)=2$.}
while
\begin{align}
p_{\by}(\bsfy) & = \sum_{k=1}^{4^{\Nt-1}} \! \frac{p_k}{4} \sum_{i=0}^3 p_{\by | \bx} (\bsfy | \j^i \bsfx_k) 
\label{Leiva2}
\end{align}
with $p_{\by | \bx}$ depending on $\SNR$ and $\bH$ as per (\ref{Joan}).
Elaborating on (\ref{Leiva2}),
\begin{align}
p_{\by}(\bsfy) & = \sum_{k=1}^{4^{\Nt-1}} \! \frac{p_k}{4} \left[ \prod_{n=0}^{\Nr-1} Q  \! \left( \! - \Re\{\sfy_n\} \sqrt{\frac{\SNR}{\Nt}} \Re\{\bh_n \bsfx_k \} \! \right)   \right. \nonumber \\
& \quad \cdot Q  \! \left( \! - \Im\{\sfy_n\} \sqrt{\frac{\SNR}{\Nt}} \Im\{\bh_n \bsfx_k\} \! \right) \nonumber \\
& \quad + \prod_{n=0}^{\Nr-1} Q  \! \left( \!  \Re\{\sfy_n\} \sqrt{\frac{\SNR}{\Nt}} \Im\{\bh_n \bsfx_k \} \! \right)   \nonumber \\
& \quad \cdot Q  \! \left( \! - \Im\{\sfy_n\} \sqrt{\frac{\SNR}{\Nt}} \Re\{\bh_n \bsfx_k \} \! \right) \nonumber \\
& \quad + \prod_{n=0}^{\Nr-1} Q  \! \left( \!  \Re\{\sfy_n\} \sqrt{\frac{\SNR}{\Nt}} \Re\{\bh_n \bsfx_k \} \! \right)   \nonumber \\
& \quad \cdot Q  \! \left( \! \Im\{\sfy_n\} \sqrt{\frac{\SNR}{\Nt}} \Im\{\bh_n \bsfx_k \} \! \right) \nonumber \\
& \quad + \prod_{n=0}^{\Nr-1} Q  \! \left( \! - \Re\{\sfy_n\} \sqrt{\frac{\SNR}{\Nt}} \Im\{\bh_n \bsfx_k \} \! \right)   \nonumber \\
& \quad \left. \cdot Q  \! \left( \! \Im\{\sfy_n\} \sqrt{\frac{\SNR}{\Nt}} \Re\{\bh_n \bsfx_k \} \! \right) \right] .
\label{MalTemps2}
\end{align}
In turn, because of the factorization of $p_{\by|\bx}$ in (\ref{Joan}),
\begin{align}
  \cH(\by | \bx) & = \! \sum_{n=0}^{\Nr-1} \big( \cH(\Re\{ y_n \} | \bx) +  \cH(\Im\{ y_n \} | \bx) \big) \\
& = \!\! \sum_{k=1}^{4^{\Nt-1}} \!  \frac{p_k}{4} \sum_{i=0}^3 \! \sum_{n=0}^{\Nr-1} \big( \cH(\Re\{ y_n \} | \bx=\j^i\bsfx_k) \nonumber \\
 & \quad +  \cH(\Im\{ y_n \} | \bx = \j^i \bsfx_k) \big) \\
 & = \!\! \sum_{k=1}^{4^{\Nt-1}} \! \frac{p_k}{4} \sum_{i=0}^3 \! \sum_{n=0}^{\Nr-1} \! \left[ \cH_{\rm b} \! \left( Q  \! \left( \! -  {\textstyle \sqrt{\frac{\SNR}{\Nt} }  } \Re\{\bh_n \j^i \bsfx_k \} \! \right) \!  \right) \right. \nonumber \\
 & \quad \left. +  \cH_{\rm b} \! \left( Q  \! \left( \! - {\textstyle  \sqrt{\frac{\SNR}{\Nt} } } \Im\{\bh_n \j^i \bsfx_k \} \! \right) \!  \right) \right]
\end{align}
where $\cH_{\rm b}(p) = -p \log_2 p - (1-p) \log_2(1-p)$ is the binary entropy function. Since changing $i$ merely flips the sign of some of the Q-funcion arguments, and $Q(-\xi) = 1 - Q(\xi)$ such that $\cH_{\rm b}(Q(-\xi)) = \cH_{\rm b}(Q(\xi))$, it follows that
\begin{align}
 \cH(\by | \bx) & = \! \sum_{k=1}^{4^{\Nt-1}} \! p_k \sum_{n=0}^{\Nr-1} \! \left[ \cH_{\rm b} \! \left( Q  \! \left( \! -   \sqrt{\frac{\SNR}{\Nt} } \Re\{\bh_n \bsfx_k \} \! \right) \!  \right) \right. \nonumber \\
 & \quad \left. +  \cH_{\rm b} \! \left( Q  \! \left( \! -  \sqrt{\frac{\SNR}{\Nt}  } \Im\{\bh_n \bsfx_k \} \! \right) \!  \right) \right] .
 \label{TrumpOut2}
\end{align}

The combination of (\ref{NoWiFi2}), (\ref{MalTemps2}), and (\ref{TrumpOut2}) gives $\cI(\SNR,\bH)$, whose evaluation involves $\cO(4^{\Nt-1} 4^{\Nr-1})$ terms. This becomes prohibitive even for modest $\Nt$ and $\Nr$, hence the interest in analytical characterizations. From $\cI(\SNR,\bH)$, the 1-bit capacity is
\be
C(\SNR,\bH) = \max_{ \{p_k \} : \sum_k p_k = 1 } \cI(\SNR,\bH)
\label{VAB}
\ee
with maximization over $p_1,\ldots,p_{4^{\Nt-1}}$.
Since $\cI(\SNR,\bH)$ is concave in $p_1,\ldots,p_{4^{\Nt-1}}$ and these probabilities define a convex set, (\ref{VAB}) can be solved with off-the-shelf convex optimization tools. Or, the Blahut-Arimoto
algorithm that alternatively maximizes $p_\bx$ and $p_{\bx|\by}$
can be applied, with converge guarantees to any desired accuracy \cite{blahut1972computation,arimoto1972algorithm}.

In ergodic settings, what applies is the ergodic spectral efficiency
\be
\cI(\SNR) = \E_\bH \big[ \cI(\SNR,\bH) \big] 
\label{RH}
\ee
and likewise for the ergodic capacity.
Alternatively, if the channel is stable over each codeword, then $\cI(\SNR,\bH)$ and $C(\SNR,\bH)$ themselves are meaningful for each $\bH$.
%$C(\SNR)$ should be interpreted as the average 1-bit capacity over the settings described by the distribution of $\bH$.

The 1-bit capacity cannot exceed $2 \min(\Nt,\Nr)$ b/s/Hz, with three distinct regimes:
\begin{itemize}
\item Low SNR. This is a key regime at mmWave and terahertz frequencies, given the difficulty in producing strong signals, the high propagation losses, and the noise bandwidth.
\item Intermediate SNR. Here, the spectral efficiency improves sustainedly with the SNR.
\item High SNR. This is a regime of diminishing returns, once the capacity nears $2 \min(\Nt,\Nr)$.
\end{itemize}

\subsection{Low SNR}

The low-SNR behavior is most conveniently examined with the mutual information expressed as function of the normalized energy per bit at the receiver,
\be
\label{ebnodef}
\ebno = \frac{\SNR}{\cI(\SNR)} .
\ee
Beyond the minimum required value of
\begin{align}
\label{eooo}
\ebnomin = \lim_{\SNR \to 0} \, \frac{\SNR}{\cI(\SNR)} ,
\end{align}
the mutual information behaves as \cite[sec. 4.2]{Foundations:18}
\be
\label{fumfumfum}
S_0 \, \frac{\left. \ebno \right |_{\rm \scriptscriptstyle dB} - \left. \ebnomin \right |_{\rm \scriptscriptstyle dB}}{3 \, {\rm \scriptstyle dB}} + \varepsilon ,
\ee
where $\varepsilon$ is a lower-order term, $S_0$ is the slope at $\ebnomininline$ in b/s/Hz/($3$ dB), and $z \dB = 10 \log_{10} z$.

\begin{figure*}
\begin{align}
S_0 = \frac{2 \, [\dot{\cI}(0)]^2}{-\ddot{\cI}(0) \log_2 e} = \frac{ \E \big[ \tr(\bH \bSigma_\bx \bH^*) \big]^2 }{\frac{1}{2} \, \E \big[ \tr \big( ( \text{nondiag}(\bH \bSigma_\bx \bH^*)   )^2  \big) \big] + \frac{\pi-1}{3} \, \E \big[ \| \Re \{ \bH \bx \} \|^4_4 + \| \Im \{ \bH \bx \} \|^4_4 \big] } 
\label{ForzaOriol}
\end{align}
\end{figure*}

$\ebnomin$ and $S_0$ descend from the first and second derivatives of $\cI(\SNR)$ at $\SNR=0$, which themselves emerge from (\ref{NoWiFi2}), (\ref{MalTemps2}), and (\ref{TrumpOut2}) after a tedious derivation \cite{mezghani2007ultra,mezghani2020low}. Plugging these two derivatives into the definitions of $\ebnomin$ and $S_0$ \cite[sec. 4.2]{Foundations:18},
\begin{align}
\ebnomin & = \frac{1}{\dot{\cI}(0)} = \frac{\pi \Nt}{ \E \big[ \tr( \bH \bSigma_\bx \bH^*  )  \big] \log_2 e }
\label{Argimon}
\end{align}
%A tedious expansion of the mutual information in (\ref{NoWiFi2}) yields \cite{mezghani2007ultra} 
%\begin{align}
%\cI(\SNR,\bH) & = \frac{1}{\pi \Nt} \, \tr \big( \bH  \bSigma_\bx  \bH^* \big) \, \SNR \, \log_2 e \nonumber \\
%& \quad - \frac{1}{\pi^2 N^2_{\rm t}} \left[  \frac{1}{2 } \, \tr \Big( \! \big( \text{nondiag}\big( \bH  \bSigma_\bx  \bH^* \big) \big)^2  \Big)  \right. \nonumber \\
%& \quad \left. + \frac{\pi-1}{3 } \, \E_\bx \big[ \| \bH \bx  \|^4_4  \big] \right] \SNR^2 \, \log_2 e \nonumber \\
%& \quad + o(\SNR^2) \label{RectorCasals}
%\end{align}
with $\bSigma_\bx = \E \big[ \bx \bx^* \big] = \sum_k p_k \, \bsfx_{k} \bsfx^*_{k} $, while $S_0$ equals (\ref{ForzaOriol})
where $\text{nondiag}(\cdot)$ returns a matrix with its diagonal entries set to zero and $\| \cdot \|_4$ denotes L4 norm.
%whereas, for an $N$-dimensional vector, $\|  \bz \|^4_4 = \sum_{j=1}^N \big( \Re\{z_j\}^4 + \Im\{z_j\}^4 \big)$.

The expectations in (\ref{Argimon}) and (\ref{ForzaOriol}) are conditioned on $\bH$ when there is operational significance attached to a specific such value, and unconditioned in the ergodic case.

\begin{figure}
	\centering
	\includegraphics[width=1\linewidth]{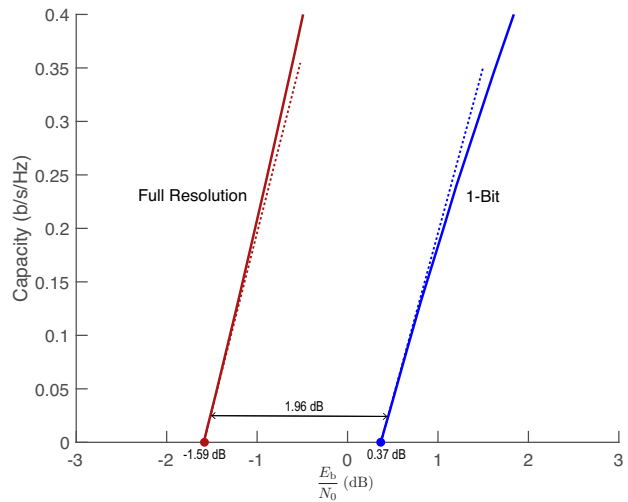}
	\caption{Capacity as a function of $E_{\rm b}/N_0$ for $\Nt=\Nr=1$ and Rayleigh fading. The solid lines are the exact capacities (1-bit and full resolution) while the dotted lines represent their respective expansions as per (\ref{fumfumfum}).}
	\label{Expansions}
\end{figure}

A worthwhile exercise is to appraise the expansion in (\ref{fumfumfum}) against its exact counterpart, a contrast that Fig. \ref{Expansions} presents for $\Nt=\Nr=1$ in Rayleigh fading. The characterization provided by (\ref{fumfumfum}) is indeed precise, a fact that extends to all other channels considered in the paper.

\subsection{Intermediate SNR}

In 1-bit communication, the intermediate-SNR regime steals relevance from its high-SNR counterpart, which becomes unappealing.
In order to delineate the reach of this intermediate-SNR regime, it is of interest to establish the limiting capacity for $\SNR \to \infty$.
Let us define
\be
C_{\infty}(\bH) = \lim_{\SNR \to \infty} C(\SNR,\bH) 
\ee
and consider channels satisfying $\bh_n \bx \neq 0$ with probability 1 for $n=0,\ldots,\Nr-1$, such that
the transition probabilities have a positive mass only at $0$ and $1$, meaning that $\by$ is fully determined by $\bx$.
The vast majority of channels abide by the condition, and in particular the ones set forth in Sec. \ref{Carla18}.

For $\Nt=1$, a single quartet is available for transmission and, by virtue of its four equiprobable constituent vectors, $C_{\infty}(\bH) = 2$.
For $\Nt>1$ and $\Nr=1$, it can be verified that (\ref{VAB}) is maximized when a single quartet is activated, depending on $\bH$ \cite{nam2019capacity}. Again, $C_{\infty}(\bH) = 2$.

For $\Nt>1$ and $\Nr>1$, it must hold that $C_\infty \leq 2 \Nt$, but this bound is generally not achievable because some vectors $\bx$ map to the same receive vector $\by$ \cite{gao2017power}.
% because of the receiver's inability to distinguish all $4^{\Nt}$ possible transmit vectors $\bx$, as some of these map to the same $\by$ \cite{gao2017power}. Indeed, any two vectors $\bsfx_1$ and $\bsfx_2$ satisfying $\text{sgn}(\bH \bsfx_1) = \text{sgn}(\bH \bsfx_2)$ cannot be distinguished.
As the transition probabilities are either $0$ or $1$, every binary entropy function in (\ref{TrumpOut2}) vanishes and $\cH(\by | \bx) \to 0$, hence the mutual information comes to equal $\cH(\by)$.
Letting 
\be
\cY(\bH) = \Big\{ \by  \, | \, \by = \text{sgn}(\bH \bx) \, \forall \bx \in \{ \pm1 \pm \j \}^{\Nr} \Big\} 
\label{Tian}
\ee
denote the set of vectors $\by$ that can be elicited for channel $\bH$, the maximization of $\cH(\by)$ occurs when this set is equiprobable. Then,
\be
C_\infty(\bH) = \log_2 |\cY(\bH)| 
\label{Tian2}
\ee 
with $\E[C_\infty(\bH)]$ being the limiting ergodic capacity. 

The evaluation of (\ref{Tian2}) is far simpler than that of $C(\SNR)$ in its full generality.

\section{1-Bit vs Full-Resolution Capacity}
\label{calor3}

A naïve comparison of the 1-bit and full-resolution capacities would indicate that the former always trails the latter.
In terms of power, their gap in dB is at least the difference between their $\ebnomin \dB$ values; in a scalar Rayleigh-faded channel, for instance, what separates the full-resolution mark of $-1.59$ dB \cite[sec. 4.2]{Foundations:18} from its 1-bit brethren of $0.37$ dB is $1.96$ dB as noted in Fig.~\ref{Expansions}. As shall be seen in the sequel, this gap remains rather steady with MIMO and over a variety of channels.

Such naïve comparison, however, only accounts for radiated power, disregarding any other differences in power consumption between the full-resolution and 1-bit alternatives. While appropriate when the radiated power dominates, this neglect becomes misleading when the digitalization consumes sizeable power and, since this is the chief motivation for 1-bit communication, by definition the comparison is somewhat deceptive.
% A more comprehensive comparison would have to entail, in lieu of the received energy per bit, $E_{\rm b}$, the total energy per bit consumed at the transmitter. 

Indeed, whenever the excess power of a full-resolution architecture, relative to 1-bit, exceeds a $1.96$-dB backoff in $\Pt / \eta$,
there is going to be a range of SNRs over which, under a holistic accounting of power, the 1-bit capacity is actually higher. % than its full-resolution brethren.
For a very conservative assessment of this phenomenon, let us assume that $\kappa=2$ in (\ref{LMessi}) and that $\eta$, $\mathsf{FoM}$, and $P_{\sf other}$, are not affected by the resolution---in actuality all of these quantities shall be markedly better in the 1-bit case---to obtain the condition
\be
\frac{1}{10^{1.96/10}} \frac{\Pt}{\eta} + \mathsf{FoM} \, B \, 2^{11} \geq \frac{\Pt}{\eta} + \mathsf{FoM} \, B \, 4
\ee
where we considered two ADCs ($\Nr=1$) and $b=10$ bits for full resolution. The above yields
\be
B \geq \frac{0.36 \, \Pt / \eta}{ \mathsf{FoM} \, (2^{11}-4)} ,
\ee
which, for the sensible values $\Pt=23$ dBm and $\eta=0.4$, and with a state-of-the-art $\mathsf{FoM}=10$ pJ/conversion \cite{murmann2016adc}, evaluates to $B \geq 8.8$ GHz. This highly conservative threshold drops rapidly as the number of digitally processed antennas grows large and thus, for bandwidths well within the scope of upcoming wireless systems, 1-bit MIMO can be viewed as information-theoretically optimum for at least some range of SNRs.

%$65 \geq \mathsf{FoM} \geq 5$ fJ/comparison according to Murmann
%
%SUNDEEP'S PAPER (RX ONLY):
%f=140 GHz
%LNA: 5 mW
%LO + mixer: 100 mW
%Filters: 13 mW
%Total: 120 mW
%CHANNEL CODER AND DECODER MISSING

\section{Transmit Beamforming}
\label{calor4}

Transmit beamforming corresponds to $\bSigma_\bx$ being $\mbox{\text{rank-1}}$, i.e., to $\bx$ being drawn from a single quartet, with such quartet generally dependent on $\bH$. We examine this strategy with an ergodic perspective; for nonergodic channels, the formulation stands without the expectations over $\bH$.

\subsection{Low SNR}

For vanishing SNR, transmit beamforming is not only conceptually appealing, but information-theoretically optimum.
Indeed, (\ref{Argimon}) can be rewritten as
\begin{align}
\ebnomin = \frac{\pi \Nt}{ \E \big[ \sum_{k} p_k  \, \| \bH \bsfx_k \|^2 \big] \log_2 e } ,
\label{Argimon2}
\end{align}
which is maximized by assigning probability $1$ to the quartet $k^\star = \arg\max \| \bH \bsfx_k \|^2$ for each realization of $\bH$.
Therefore, it is optimum to beamform, and the optimum beamforming quartet is the one maximizing the received power.
%The asymptotically optimum strategy at low SNR is thus to activate a single channel-dependent quartet.
%$\bSigma_\bx$ is then rank-1, which is the broad definition of beamforming.
%(If $k^\star$ is multiple, then $\ebnomininline$ does not depend on how the transmit probabilities are split among those quartets. Activating a single one continues to be optimum from that standpoint, while activating all of them uniformly extends the optimality to $S_0$.)
%This parallels the optimum strategy with full-resolution converters, namely beamforming. Indeed, full-resolution beamforming amounts to modulating the value of $\bx$ that maximizes $\| \bH \bx \|^2$, i.e., the maximum-eigenvalue eigenvector of $\bH^* \bH$. %The 1-bit beamforming transmission is embodied by phase-modulated versions of $\bsfx_{k^\star}$.
The task is then to determine $k^\star$ from within the $4^{\Nt-1}$ possible quartets.
%The dual challenge posed by beamforming is to determine $k^\star$ from within the $4^{\Nt-1}$ possible quartets and to compute the ensuing $\ebnomininline$.

For $\Nt=1$, there is no need to optimize over $k$---only one quartet can be transmitted---and thus
\begin{align}
\ebnomin %& = \frac{\pi}{2 \, \E \! \left[  \| \bh \|^2 \right]  \log_2 e}  \\
& = \frac{\pi }{2 \Nr \log_2 e} ,
\end{align}
which amounts to $0.37$ dB for $\Nr=1$ \cite{Verdu2002} and improves by $3$ dB with every doubling of $\Nr$ thereafter.

For $\Nt > 1$, it is useful to recognize that
%maximizing $ \| \bH \bsfx_k \|$ is tantamount to maximizing the minimum distance among (the projections at the receiver of) the $k$th quartet vectors at the receiver \cite{}. Intuitively,
the choices for $\bx$ that are bound to yield high values for $\| \bH \bx \|^2$ are those that project maximally on the dimension of $\bH$ that offers the largest gain, namely the maximum-eigenvalue eigenvector of $\bH^* \bH$. This, in turn, requires that $\bx$ mimic, as best as possible, the structure of that eigenvector; since the magnitude of the entries of $\bx$ is fixed, this mimicking ought to be in terms of phases only.
Formalizing this intuition, it is possible to circumvent the need to exhaustively search the entire field of $4^{\Nt-1}$ possibilities and
 conveniently identify a subset of only $\Nt$ quartet candidates that is sure to contain the one best aligning with the maximum-eigenvalue eigenvector of $\bH^* \bH$, denoted henceforth by $\bv_0$.
Precisely, as detailed in Appendix \ref{superlliga}, if we let $\varphi_m = \angle(v_{0,m}) + \epsilon$ for $m=0,\ldots,\Nt-1$, the $\Nt$ quartets in the subset can be determined as
\be 
\bx_k = \text{sgn} \big( e^{\j \varphi_{k-1}} \bv_0 \big) \qquad\quad k=1,\ldots,\Nt
\label{subset}
\ee 
where $\epsilon$ is a small quantity, positive or negative. If the channel is rank-1, then this subset is sure to contain the optimum $\bx_{k^\star}$; if the rank is higher, then optimality is not guaranteed, but the best value in the above subset is bound to yield excellent performance.

Turning to the $\ebnomininline$ achieved by $\bx_{k^\star}$, its explicit evaluation is complicated, yet its value can be shown (see Appendix \ref{superlliga} again) to satisfy
\begin{align}
%\frac{\pi \Nt}{2  \sum_{m=1}^{\Nt}  \E \big[  \lambda_m |  \bv^*_m \, \bsfv_{1}  |^2  \big] \log_2 e} & \leq
\frac{\pi}{2 \, \E \big[ \lambda_0 \big] \log_2 e}    \leq  \ebnomin  \leq \frac{\pi^3 \Nt}{16 \, \E \big[  \lambda_{0}  \| \bv_{0} \|^2_1 \big] \log_2 e}
\label{bounds}
\end{align}
where $\lambda_0$ is the maximum eigenvalue of $\bH^* \bH$
while $\|  \cdot \|_1$ denotes L1 norm.
% and $\bsfv_1$ has unit-magnitude entries with $\angle(\sfv_{1,m})=\angle(v_{1,m})$ for $m=1,\ldots,\Nt$.
% whose phases are those of the entries of $\bv_1$.
% $[\bsfv_1]_n = [\bv_1]_n/|  [\bv_1]_n |$ 
For $\Nr=1$, (\ref{bounds}) specializes to
\begin{align}
%\frac{\pi }{2  \left(1+(\Nt-1) \frac{\pi}{4} \right)  \log_2 e} & \leq
\frac{\pi}{2 \Nt \log_2 e}    \leq  \ebnomin  \leq \frac{\pi^3}{16    \left(1+(\Nt-1) \frac{\pi}{4} \right)   \log_2 e} .
\nonumber
\end{align}

Finally, $S_0$ can be obtained by plugging $\bx = \bx_{k^\star}$ and $\bSigma_\bx = \bx_{k^\star} \bx^*_{k^\star}$ into (\ref{ForzaOriol}).

\subsection{Intermediate SNR}

The low-SNR linearity of the mutual information in the received power is the root cause of the optimality of power-based beamforming in that regime. The orientation on the complex plane of the received signals is immaterial---a rotation shifts power from the real to the imaginary part, or vice versa, but the total power is preserved. Likewise, the power split among receive antennas is immaterial to the low-SNR mutual information.

At higher SNRs, the linearity breaks down and the mutual information becomes a more intricate function of $\bH \bx$, such that proper signal orientations and power balances become important, to keep $\bh_n \bx$ away from the ADC quantization boundaries for $n=0,\ldots,\Nr-1$. This has a dual consequence: 
\begin{itemize}
\item Transmit beamforming ceases to be generally optimum, even if the channel is rank-1.
\item Even within the confines of beamforming, solutions not based on maximizing power are more satisfying.
\end{itemize}

\begin{figure}
	\centering
	\includegraphics[width=1\linewidth]{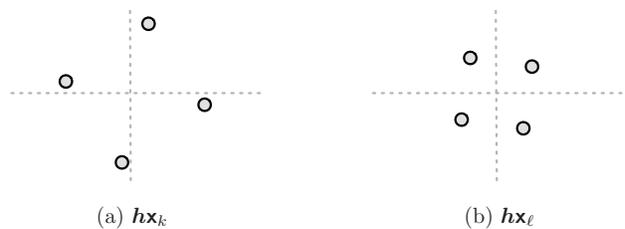}
	\caption{Complex plane representation of the four values of $\bh \bx$ for a given $\bh$ and a given quartet, with the ADC quantization boundaries indicated by dashed lines. Left-hand side, for $\bsfx_{k}$, which has a larger magnitude but worse orientation. Right-hand side, for $\bsfx_{\ell}$, which has a smaller magnitude but better orientation. On this channel, quartet $k$ yields a higher mutual at low SNR while quartet $\ell$ yields a higher mutual information beyond the low-SNR regime.}
	\label{comparison}
\end{figure}

As exemplified in Fig. \ref{comparison} for $\Nr=1$, a beamforming quartet with a better complex-plane disposition at the receiver may be preferable to one yielding a larger magnitude.
This is because, after a 1-bit ADC, only $90^\circ$ rotations and no scalings are possible (in contrast with full-resolution receivers, where $\bh \bx$ can subsequently be rotated and scaled). The best beamforming quartet is the one that simultaneously ensures large real and imaginary parts for $\bh_n \bx$ in a balanced fashion for $n=0,\ldots,\Nr-1$, and the task of identifying this quartet is a fitting one for learning algorithms \cite{9210010,Eusipco21}.

%Although it may be tempting to think that the optimum beamforming strategy is equivalent to selecting the quartet that best align with $\bh$, this is not the case. Alignment with $\bh$ maximizes the magnitude of $\bh \bx$, but it does not optimize its orientation with respect to the quantization boundaries imposed by the ADC. With a full resolution ADC, $\bh \bx$ could be rotated and scaled at will by the receiver, hence the only design criterion for the transmit beamforming would be to maximize its magnitude, prior to noise addition. With a 1-bit ADC, however, only $90^\circ$ rotations and no scalings are possible after digitalization, hence the design criterion for the beamforming generalizes to finding the quartet that yields the points $\bh \bx$ best positioned on the complex plane:
%\begin{itemize}
%\item A large magnitude for $\bh \bx$ is desirable, as it pushes the signal away from the origin, where it is highly vulnerable to noise.
%\item A proper orientation is also important, to keep $\bh \bx$ away from the quantization boundaries. As exemplified in Fig. \ref{comparison}, a better orientation may be preferable even at the expense of a smaller magnitude.
%\end{itemize} 

We note that, with full-resolution converters, multiple receive antennas play a role dual to that of transmit beamforming \cite[sec. 5.3]{Foundations:18}, and the spectral efficiency with $N$ transmit and one receive antenna equals its brethren with one transit and $N$ receive antennas. With 1-bit converters, in contrast, transmit beamforming optimizes $\bh_n \bx$ for $n=0,\ldots,\Nr-1$, to mitigate the addition of noise prior to quantization, while multiple receive antennas yield a diversity of quantized observations from which better decisions can be made on which of the possible vectors was transmitted. This includes majority decisions and erasure declarations in the case of split observations.

\section{Equiprobable Signaling}
\label{calor5}

The complementary strategy to beamforming is to activate multiple quartets, increasing the rank of $\bSigma_\bx$.
Ultimately, all quartets can be activated with equal probability, such that $\bSigma_\bx = 2 \bI$.
This renders the signals IID across the transmit antennas, i.e., pure spatial multiplexing.
We examine this strategy with an ergodic perspective.

\subsection{Low SNR}

With equiprobable signaling, (\ref{Argimon}) gives
\be
\ebnomin =\frac{\pi}{2 \Nr \log_2 e}  .
\label{Zaira}
\ee
In addition \cite{mezghani2007ultra},
\begin{align}
\E_\bx \big[ \| & \Re \{ \bH  \bx \} \|^4_4  + \| \Im \{ \bH \bx \} \|^4_4 \big]  = 6 \, \tr \big( ( \text{diag}(\bH \bH^*) )^2  \big) \nonumber  \\
& - 4 \sum_{n=0}^{\Nr-1} \sum_{m=0}^{\Nt-1} \left(\Re\{h_{n,m}\}^4 + \Im\{h_{n,m}\}^4 \right) ,
\label{Nil2}
\end{align}
based on which $S_0$ in (\ref{ForzaOriol}) simplifies considerably.

Combining (\ref{bounds}) and (\ref{Zaira}), the low-SNR advantage of optimum beamforming over equiprobable signaling, denoted by $\Delta_{\sf BF}$, is tightly bounded as
\be
 \frac{8 \, \E \big[  \lambda_{0}  \| \bv_{0} \|^2_1 \big]}{\pi^2 \Nt \Nr}  \leq \Delta_{\sf BF} \leq \frac{\E \big[ \lambda_0 \big]}{\Nr}  .
 \label{Oscar}
\ee
%For its part, the shortfall of 1-bit optimum beamforming relative to full-resolution beamforming, $\Delta_{\sf 1bit}$, lies within
%\be
%1.96 \, \text{dB} \leq \Delta_{\sf 1bit} \leq  \frac{\pi^3 \Nt \, \E[\lambda_1]}{16 \, \E \big[  \lambda_{1}  \| \bv_{1} \|^2_1 \big]} \dB .
%\label{saw}
%\ee
This enables some general considerations:
\begin{itemize}
\item The low-SNR advantage of beamforming is essentially determined by the maximum eigenvalue of $\bH^*\bH$. The advantage is largest in rank-1 channels, and minimal if all eigenvalues are equal (on average or instantaneously, as pertains to ergodic and nonergodic settings).
\item If all eigenvalues are equal, beamforming may still yield a lingering advantage for $\Nt>\Nr$, but not otherwise. Indeed, for $\Nt \leq \Nr$, if all eigenvalues all equal then $\E \big[ \lambda_0 \big] = \Nr$ and thus $\Delta_{\sf BF} \leq 1$.
\end{itemize}

\subsection{Intermediate SNR}

While beamforming is optimum at low SNR, it is decidedly suboptimum beyond, and activating multiple quartets becomes instrumental to surpass the 2-b/s/Hz mark. This is the case even in rank-1 channels, where the activation of multiple quartets allows producing richer signals; this can be seen as the 1-bit counterpart to higher-order constellations. And, given how the curse of dimensionality afflicts the computation of the optimum quartet probabilities, equiprobable signaling is a very enticing way of going about this.
As will be seen, not only is it implementationally convenient, but highly effective.
% and often superior to other alternatives, for instance quantizing the transmit signal that would be optimum under full-resolution converters, that is, the modulated eigenvectors of $\bH^*\bH$ \cite[sec. 5.3]{Foundations:18}.
%Clearly, the highly nonlinear quantization disrupts the delicate orthogonal structure of the solution intended for linear converters.

\section{Channels of Interest}
\label{calor6}

Capitalizing on the analytical tools set forth hitherto, let us now examine the performance of transmit beamforming and equiprobable signaling in various classes of channels, starting with the nonergodic LOS settings and progressing on to the ergodic IID Rayleigh-faded channel.

\subsection{LOS with Planar Wavefronts}

\begin{figure}
	\centering
	\includegraphics[width=1\linewidth]{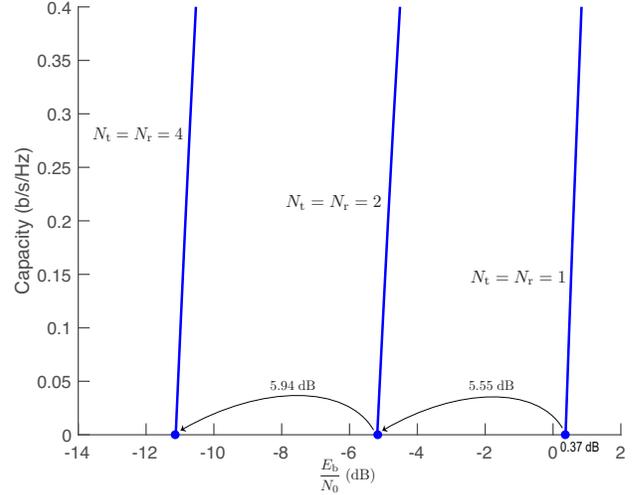}
	\caption{Capacity as a function of $E_{\rm b}/N_0$ for $\Nt=\Nr=1$, $\Nt=\Nr=2$, and $\Nt=\Nr=4$, in a planar-wavefront LOS channel with half-wavelength antenna spacings, $\theta_{\rm t}=0$, $\theta_{\rm r}=\pi/6$, and $\phi=\pi/4$. %In solid, exact values; shaded, the intervals spanned by the combination of (\ref{noies1}) and (\ref{noies2}).
	}
	\label{CapMIMObfRank1}
\end{figure}

This channel is rank-1, hence the optimum $\ebnomininline$ can be achieved with equality by the best beamforming quartet in subset (\ref{subset}).
More conveniently for our purposes here, we can rewrite (\ref{Oscarinyu}) as $\bH = \sqrt{\Nt \Nr} \bu \, \bv^*$ where
\begin{align}
u_n & = \frac{1}{\sqrt{\Nr}} \, e^{-j \pi \frac{2n}{\lambda} \dr \sin \! \thetar \cos \! \phi  } \\
v_m & = \frac{1}{\sqrt{\Nt}} \, e^{-j \pi  \frac{2m}{\lambda} \dt \sin \! \thetat }. 
\end{align}
%the entries of $\bu$ and $\bv$ have, respectively, magnitudes $1/\sqrt{\Nr}$ and $1/\sqrt{\Nt}$, and phases as (\ref{ventvent}).
%the transmit quartet that maximizes $\| \bH \bx \|^2$.
Irrespective of the array orientations, $\lambda_0 = \Nt \Nr$ and $\| \bv_0 \|^2_1 = \Nt$ such that (\ref{bounds}) reverts to
\begin{align}
\frac{\pi}{2 \, \Nt \Nr \log_2 e}    \leq  \ebnomin  \leq \frac{\pi^3 }{16 \, \Nt \Nr \log_2 e}  ,
\label{noies1}
\end{align}
which depends symmetrically on $\Nt$ and $\Nr$.
%In turn, as shown in Appendix \ref{Noah},
%\be
%\frac{\frac{128}{\pi^4}  \Nr}{\Nr + \frac{ 2 \pi -5}{3}} \leq S_0 \leq  \frac{\frac{\pi^4}{32} \Nr}{\Nr + \frac{ 2 \pi -5}{3}} ,
%\label{noies2}
%\ee
%which depends only mildly on the antenna counts, such that the low-SNR performance is dictated essentially by $\ebnomininline$.
The significance of $\ebnomin$ as the key measure of low-SNR performance can be appreciated in Fig. \ref{CapMIMObfRank1}, which depicts the low-SNR capacity as a function of $\ebnoinline$ for $\Nt=\Nr=1$, $2$, and $4$ in an exemplary LOS setting. Adding antennas essentially displaces the capacity by the amount by which $\ebnomin$ changes.

\begin{figure}
	\centering
	\includegraphics[width=1\linewidth]{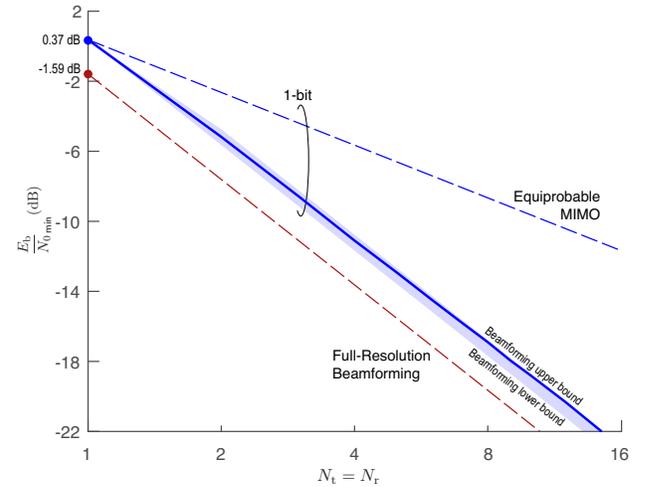}
	\caption{Minimum $E_{\rm b}/N_0$ as a function of $\Nt=\Nr$ for a planar-wavefront LOS channel with half-wavelength antenna spacings, $\theta_{\rm t}=0$, $\theta_{\rm r}=\pi/6$, and $\phi=\pi/4$: 1-bit beamforming (exact values in solid, interval spanned by the bounds in shaded) vs equiprobable signaling. Also shown is the performance with full resolution.}
	\label{EbN0min_vs_N_THz}
\end{figure}

Shown in Fig. \ref{EbN0min_vs_N_THz} is how $\ebnomininline$ improves with the number of antennas ($\Nt=\Nr$) for the same setting. Also shown are the values for equiprobable signaling, undesirable in this case as per (\ref{Oscar}). %and the values for full-resolution beamforming.
The low-SNR advantage of beamforming accrues steadily with the numbers of antennas and the bounds in (\ref{bounds}) tightly bracket the optimum $\ebnomininline$. As anticipated, the gap of 1-bit beamforming to full-resolution beamforming (included in the figure) remains small.

%From (\ref{saw}), the 1-bit shortfall satisfies $\Delta_{\sf 1bit} \leq 2.87$ dB, a very modest deficit that is reflected in the figure.

\begin{figure}
	\centering
	\includegraphics[width=1\linewidth]{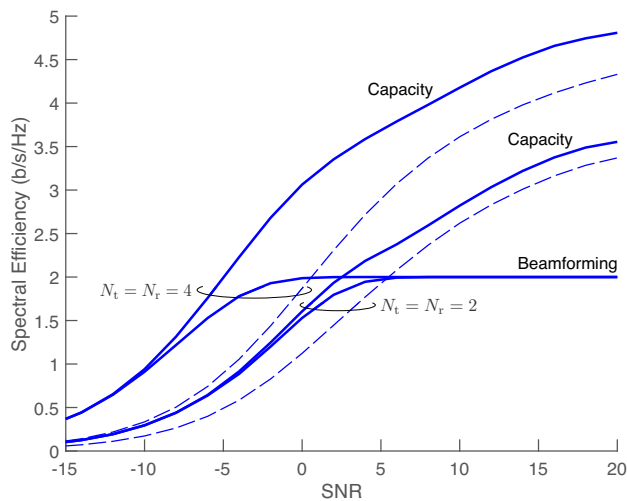}
	\caption{Spectral efficiency as a function of $\SNR$ for $\Nt=\Nr=2$ and $\Nt=\Nr=4$, in a planar-wavefront LOS channel with half-wavelength antenna spacings, $\theta_{\rm t}=\pi/4$, $\theta_{\rm r}=\pi/6$, and $\phi=\pi/4$. In solid, capacity and beamforming; in dashed, equiprobable signaling.}
	\label{CvsSNR_LOS_planar}
\end{figure}

Moving up to intermediate SNRs, the beamforming and equiprobable-signaling performance on another setting is presented in Fig. \ref{CvsSNR_LOS_planar}. 
Also shown is the actual capacity with $p_1,\ldots,p_{4^{\Nt-1}}$ optimized via Blahut-Arimoto. Up to when the $2$-b/s/Hz ceiling is approached,
beamforming performs splendidly. Past that level, and no matter the rank-1 nature of the channel, equiprobable signaling is highly superior, tracking the capacity to  within a roughly constant shortfall.
%By activating multiple subsets, the transmitter can essentially project the equivalent of higher-order constellations on the channel's eigendirection, thereby overcoming the $2$-b/s/Hz limitation.
This example represents well the intermediate-SNR performance in planar-wavefront LOS channels, a point that has be verified by contrasting the asymptotic performance of equiprobable signaling in a variety of such channels against the respective $C_\infty$.

\subsection{LOS with Spherical Wavefronts}

The scope of channels in this class is very large, depending on the array topologies and relative orientations; for the sake of specificity, we concentrate on ULAs, and draw insights whose generalization would be welcome follow-up work.
A key property of ULA-spawned channels within this class is that \cite{do2020reconfigurable}
\be
\bH^* \bH \approx \frac{\Nmax}{\eta} \bD^*_{\rm tx} \bF \text{diag}( \underbrace{1,\ldots,1}_{\eta \Nmin} , 0,\ldots,0 ) \bF^*  \bD_{\rm tx} \nonumber
\ee
%BEAMFORMING CAN TAKE PLACE ON ANY OF THE LEADING $\Nmin$ COLUMNS OF $\bF$, SAY THE FIRST ONE WHOSE ENTRIES ARE ALL $1/\sqrt{\Nt}$. THE OPTIMUM TX QUARTET HAS THEN ALL ENTRIES EQUAL TO $1+\j$ AND THE OPTIMUM LOWER BOUND ON $\ebnomin$ IS ACHIEVED.
%THE SAME $\ebnomin$ IS OBTAINED WITH EQUIPROBABLE SIGNALING OVER THE LEADING $\Nmin$ COLUMNS OF $\bF$, AND WITH EQUIPROBABLE SIGNALING FOR $\Nt \leq \Nr$.
where the approximation sharpens with the numbers of antennas while $\bF$ is a unitary Fourier matrix, $ \bD_{\rm tx}$ and $\eta$ are as introduced in Sec. \ref{Carla18}, and $\Nmin = \min(\Nt,\Nr)$. Therefore, $\lambda_0 \approx \Nmax/\eta$ and $\| \bv_0 \|^2_1 \approx \Nt$.

By specializing (\ref{bounds}), the optimum $\ebnomininline$ attained by beamforming is seen to satisfy
\begin{align}
\frac{\pi \eta}{2 \Nmax \log_2 e}    \lesssim     \ebnomin  \lesssim \frac{\pi^3 \eta }{16  \Nmax \log_2 e}  ,
\label{SIlla}
\end{align}
which indicates that
%\be
%\frac{\frac{128}{\pi^4}  \Nr}{\Nr + \frac{ 2 \pi -5}{3}} \lesssim S_0 \lesssim  \frac{\frac{\pi^4}{32} \Nr}{\Nr + \frac{ 2 \pi -5}{3}} .
%\label{noies3}
%\ee
%$\ebnomininline$ improves with diminishing $\eta$; the conclusion is that
a smaller $\eta$ is preferable at low SNR, meaning antennas as tightly spaced as possible---this renders the wavefronts maximally planar---and array orientations as endfire as possible---this shrinks their effective widths. Indeed, wavefront curvatures trim the beamforming gains, and reducing $\eta$ mitigates the extent of such curvatures.

\begin{figure}
	\centering
	\includegraphics[width=1\linewidth]{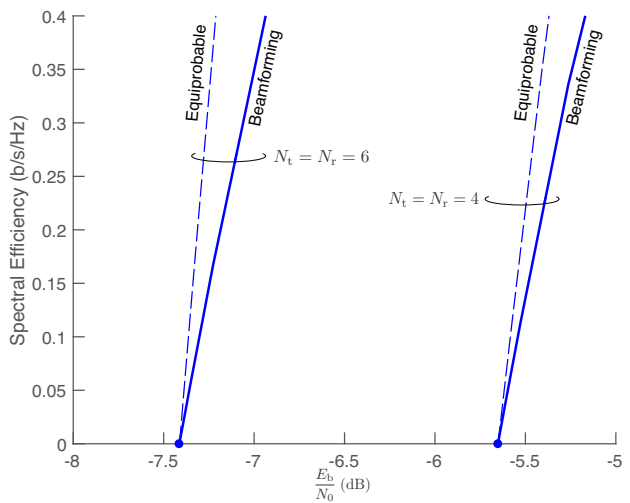}
	\caption{Spectral efficiency as a function of $E_{\rm b}/N_0$ for $\Nt=\Nr=4$ and $\Nt=\Nr=6$ in an LOS channel with $\eta=1$. In solid, beamforming performance; in dashed, capacity with equiprobable signaling.}
	\label{CapMIMObfeq}
\end{figure}

%This is as with full-resolution converters \cite{do2020reconfigurable}.
With growing $\eta$, the low-SNR performance does degrade, but beamforming retains an edge over equiprobable signaling for $\eta<1$ or $\Nt > \Nr$. Alternatively, for $\eta=1$ and $\Nt = \Nr$, (\ref{SIlla}) is no better than the equiprobable-signaling $\ebnomininline$ in (\ref{Zaira}).
In fact, for this all-important configuration whose eigenvalues are equal \cite{Haustein:03,Bohagen:05,Larsson:05,Sarris:07,Sheldon:082,do2021terahertz}, 
\emph{any} transmission strategy achieves this same $\ebnomininline$; 
%Let us now see what unfolds for a given $\eta$. Beamforming attains the optimum $\ebnomininline$, but, because $\bH$ may have various large singular values of comparable strength, other strategies may also approach such $\ebnomininline$. 
%In fact, for $\eta=1$, which is an extensively studied configuration \cite{}, 
%
%Equiprobable signaling achieves (\ref{Zaira}) regardless of $\eta$ because $\tr \big( \bH \bH^* \big) = \Nt \Nr$.
%For $\eta=1$ specifically, (\ref{Zaira}) equals the $\ebnomininline$ attained by beamforming, and in fact by every transmission strategy, and it is therefore optimum.
this can be verified by using $\bH^* \bH= \Nr \bI$ and $\| \bsfx_k \|^2 = 2 \Nt$ in (\ref{Argimon2}), whereby (\ref{Zaira}) emerges irrespective of $p_1,\ldots,p_{4^\Nt-1}$.
%To verify this, rewrite (\ref{debat}) as
%\begin{align}
%C(\SNR,\bH) & = \frac{\SNR}{\pi \Nt} \sum_{k=1}^{4^{\Nt-1}} \! p_k \, \bsfx^*_k \bH^* \bH \bsfx_k  \log_2 e \nonumber \\
%& \quad + \cO(\SNR^2) 
%\end{align}
%and, since $\bH^* \bH= \Nr \bI$ and $\| \bsfx_k \|^2 = 2 \Nt$ for every $k$, (\ref{Zaira}) emerges irrespective of $p_1,\ldots,p_{4^\Nt-1}$.
This coincidence of $\ebnomininline$ for all transmission strategies when $\eta=1$ and $\Nt = \Nr$ does not translate to $S_0$, which is decidedly larger for equiprobable signaling, indicating that this is the optimum low-SNR technique for this configuration as illustrated in Fig. \ref{CapMIMObfeq}.
Precisely, applying (\ref{ForzaOriol}) and (\ref{Nil2}), a channel with $\eta=1$ and $\Nt=\Nr=N$ is seen to exhibit
\be
S_0 = \frac{ \frac{2}{\pi-1} N^4 }{N^3-\frac{2}{3} { \displaystyle \sum_{n=0}^{N-1} \sum_{m=0}^{N-1} } \left[ \cos^4 \! \big(2 \pi \frac{n m}{N} \big) + \sin^4 \! \big(2 \pi \frac{n m}{N} \big) \right] } .
\ee

\begin{figure}
	\centering
	\includegraphics[width=1\linewidth]{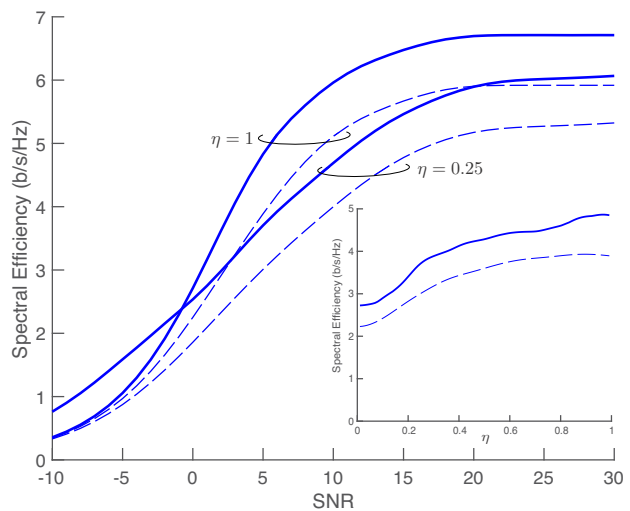}
	\caption{Main plot: spectral efficiency as a function of $\SNR$ for $\Nt=\Nr=4$, both with optimized (solid lines) and with uniform (dashed lines) quartet probabilities. Inset: spectral efficiency as a function of $\eta$ for $\SNR = 5$ dB. The channel is LOS and the arrays are broadside with $d_{\rm t} = d_{\rm r}$.}
	\label{RobNur}
\end{figure}

Let us now turn to intermediate SNRs, where the full-resolution wisdom is that the performance depends only on $\eta$ and it improves monotomically with $\eta$ up to $\eta=1$, where capacity is achieved by IID signaling.
All these insights, underpinned by the approximate equality of the $\eta \Nmin$ nonzero eigenvalues of $\bH^* \bH$, cease to hold in the 1-bit realm due to the transmitter's inability of accessing those singular values directly via precoding.
%Interference can only be managed by adjusting the quartet probabilities.
Indeed, when the only ability is to manipulate the quartet probabilities (see Fig. \ref{RobNur} for an example):
\begin{itemize}
\item The performance does not depend only on $\eta$, but further on $\theta_{\rm t}$, $\theta_{\rm r}$, $\phi$, $D$, and $d_{\rm t}$ and $d_{\rm r}$.
%\item It improves with growing $\eta$, but far less decidedly than with full-resolution converters, and may in fact be nonmonotonic. 
\item The optimum configuration need not correspond to $\eta=1$. % and the capacity-achieving signaling therein need not be equiprobable.
\end{itemize}
The main takeaway for our purpose, though, is that at intermediate SNRs equiprobable signaling closely tracks the capacity.

%For $0 < \eta <1$, the optimum signaling strategy in terms of both $\ebnomininline$ and $S_0$ might entail equiprobably activating a subset of the transmit quartets, and formalizing this by capitalizing on (\ref{nata}) is a potentially interesting line for subsequent work.

\subsection{IID Rayleigh Fading}

\begin{figure}
	\centering
	\includegraphics[width=1\linewidth]{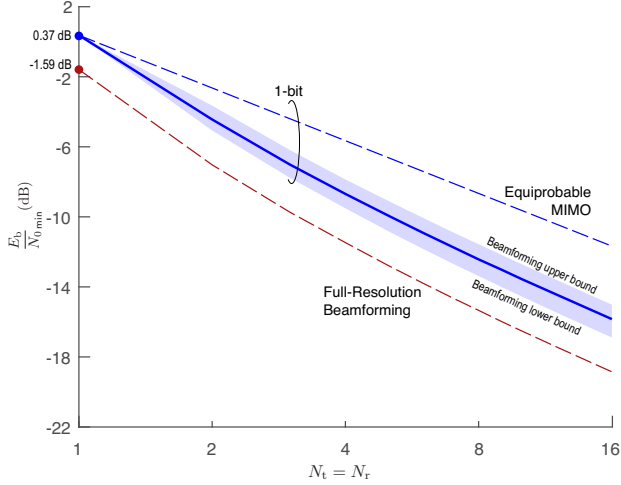}
	\caption{Minimum $\ebnoinline$ vs $\Nt=\Nr$ for an IID Rayleigh-faded channel: 1-bit beamforming (exact values in solid, interval spanned by the bounds in shaded) vs equiprobable signaling. Also shown is the performance with full resolution.}
	\label{EbN0min_vs_N}
\end{figure}

For $\bH$ having IID complex Gaussian entries, we resort to the ergodic interpretation. Shown in Fig. \ref{EbN0min_vs_N} is the evolution of $\ebnomininline$ with the number of antennas for the optimum strategy (beamforming on every channel realization) as well as for equiprobable signaling.
The bounds in (\ref{bounds}) provide an effective characterization of the optimum $\ebnomin$. Moreover, $\lambda_0$ and $\bv_0$ are independent \cite[lemma 5]{lozano2003multiple} and, although $\E[\lambda_0]$ does not lend itself to a general characterization, for growing $\Nt$ and $\Nr$ it approaches $( \sqrt{\Nt} + \sqrt{\Nr} )^2$. Thus, beamforming achieves
\begin{align}
  \frac{\pi}{2 \, \big( \sqrt{\Nt} + \sqrt{\Nr} \big)^{2} \log_2 e} & \lesssim \ebnomin \label{hamburguesada} \\
&  \lesssim \frac{\pi^3 \Nt}{16 \, \big( \sqrt{\Nt} + \sqrt{\Nr} \big)^{2} \, \E \big[ \| \bv_0 \|^2_1  \big]} \nonumber
\end{align}
which sharpens with $\Nt$ and $\Nr$. For $\Nt=\Nr=64$, for instance, (\ref{hamburguesada}) gives $\ebnomin \in [-21.77 , -23,71]$ dB, correctly placing the actual value of $-22.21$ dB. The term $\E \big[ \| \bv_0 \|^2_1  \big]$ is readily computable for given values of $\Nt$ and we note, 
%The corresponding lower bound in (\ref{}) involves $\E \big[ \| \bv_0 \|^2_1  \big]$, which appears even harder to tame; as a possible path to its characterization, we note that $\bv_0$ is a column of a standard unitary matrix, uniformly distributed over an $\Nt$-dimensional sphere.
as a possible path to taming it anaytically, that $\bv_0$ is a column of a standard unitary matrix, uniformly distributed over an $\Nt$-dimensional sphere.

%FOR IID RAYLEIGH FADING, $\lambda_1$ AND $\bv_1$ ARE INDEPENDENT \cite[Lemma 5]{lozano2003multiple}. THIS ENABLES FURTHER SIMPLIFICATIONS. ALSO, $\bV$ IS A HAAR MATRIX AND THE ENTRIES OF $\bv_1$ ARE IDENTICALLY DISTRIBUTED (SURELY NOT INDEPENDENT, AT LEAST FOR SMALL DIMENSIONALITIES). FOR LARGE DIMENSIONS, $\E[\lambda_1] \to (\sqrt{\Nt} + \sqrt{\Nr})^2$.

%Shown in Fig. \ref{MIMOvsBF2} is the ergodic spectral efficiency as a function of $\ebno$ in IID Rayleigh fading, for $\Nt=\Nr=1$ and $\Nt=\Nr=2$.
%For the latter, beamforming is seen to indeed match the optimized MIMO---not only at $\ebnomininline$ but over a considerable range---and to clearly outperform equiprobable MIMO. Eventually though, the 2-b/s/Hz limitation that beamforming is subject to does set in, as evidenced in Fig. \ref{MIMOvsBF}.

%The evolution of $\ebnomin$ with the number of antennas is illustrated in Fig. \ref{EbN0min_vs_N} for both 1-bit beamforming and full-resolution beamforming.
%Their gap is steady. The advantage of 1-bit beamforming over equiprobable MIMO can also be appreciated. DOES THIS ADVANTAGE SETTLE TO A FIXED LOSS?

With equiprobable signaling, $\ebnomin$ is given by (\ref{Zaira}) and we can further characterize $S_0$.
Starting from (\ref{ForzaOriol}) and using
\begin{align}
\big(\text{nondiag}(\bH \bH^*)\big)^2 & = (\bH\bH^*)^2 - \bH \bH^* \, \text{diag}(\bH \bH^*) \nonumber \\
& \quad - \text{diag}(\bH \bH^*) \, \bH \bH^* \nonumber \\
& \quad + \big( \text{diag}(\bH \bH^*) \big)^2
\end{align}
in conjuntion with \cite[lemma 4]{lozano2003multiple}
\begin{align}
\E \big[ \tr \big( ( \bH \bH^*)^2  \big) \big] & = \Nt \Nr \, (\Nt + \Nr) \\
\E \big[ \tr \big( \bH \bH^* \, \text{diag}(\bH \bH^*)  \big) \big] & = \Nt \Nt (\Nt+1) \\
\E \big[ \tr \big(  \text{diag}(\bH \bH^*) \, \bH \bH^* \big) \big] & = \Nt \Nt (\Nt+1) \\
\E \big[ \tr \big( ( \text{diag}(\bH \bH^*) )^2  \big) \big] & = \Nt \Nt (\Nt+1)
\end{align}
we have that
\be
\E \Big[ \tr \big( (\text{nondiag}(\bH \bH^*) )^2 \big)   \Big] = \Nt \Nt \, (\Nr - 1) .
\ee
In turn,
\be
\E \big[ \| \bH \bx \|^4_4 \big] = 6 N^2_{\rm t} \Nr
\ee
and, altogether,
\be
S_0 = \frac{2 \Nt \Nr}{ (\pi-1) \Nt + \Nr -1} ,
\ee
which is an increasing function of both $\Nt$ and $\Nr$.

%\begin{figure}
%	\centering
%	\includegraphics[width=1\linewidth]{MIMOvsBF2}
%	\caption{Ergodic spectral efficiency as a function of $\ebnomininline$ for $\Nt=\Nr=1$ and $2$ with 1-bit DACs and ADCs: MIMO with optimized and uniform quartet probabilities vs beamforming (\ref{subset}). The channel is IID Rayleigh-faded.}
%	\label{MIMOvsBF2}
%\end{figure}

\begin{figure}
	\centering
	\includegraphics[width=1\linewidth]{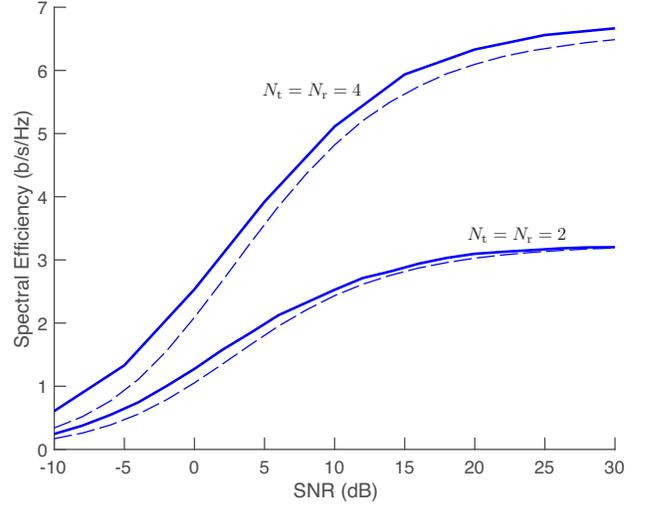}
	\caption{Ergodic spectral efficiency vs $\SNR$ for $\Nt=\Nr=2$ and $4$ both with optimized (solid lines) and with equiprobable (dashed lines) signaling. The channel is IID Rayleigh-faded.}
	\label{C1thruC4_MIMO_IID}
\end{figure}

At intermediate SNRs, equiprobable signaling is remarkably effective (see Fig.~\ref{C1thruC4_MIMO_IID}).
At the same time, the complexity of computing the mutual information---for equiprobable signaling, let alone with optimized quartet probabilities---is compounded by the need to expect it over the distribution of $\bH$, to the point of becoming unwieldy even for very small antenna counts.
Analytical characterizations are thus utterly necessary, and it is shown in Appendix \ref{Topo} that
\begin{align}
\E_\bH \big[  \cI(\SNR,\bH) \big] & = \E_\bH \big[  \cH(\by) \big] \label{Lluc1} \\
& \!\!\!\!\!\!\!\!\!\!\!\!\!\!  - \frac{2 \Nr}{\sqrt{2 \pi}} \! \int_{-\infty}^\infty \! \cH_{\rm b} \! \left( Q  \! \left( \! -   \sqrt{\SNR } \, \xi \! \right) \!  \right) e^{-\xi^2/2} \, \mathrm{d}\xi  \nonumber
\end{align}
with
\begin{align}
& 2 \Nr \geq   \E_\bH \big[  \cH(\by) \big]  \geq 2 \Nt - 2 \Nr \nonumber \\
& \qquad\;\;\,   - \log_2 \! \Bigg[ \sum_{i=0}^\Nt
 \left( \!\!
\begin{array}{c}
\Nt  \\
i 
\end{array}
\!\! \right)^{\!\! 2}
\left( \frac{1}{4 \pi^2} \arccos^2 \! \left( \frac{2 i}{\Nt} -1 \right) \! \right)^{\! \Nr} \nonumber \\
& \qquad\;\;\, + 2  \sum_{i=0}^{\Nt} \sum_{j=i+1}^{\Nt} 
\left( \!\!
\begin{array}{c}
\Nt  \\
i 
\end{array}
\!\! \right) \!
\left( \!\!
\begin{array}{c}
\Nt  \\
j 
\end{array}
\!\! \right)
 P^\Nr_\cap(i,j)
 \Bigg]   \label{XSM}
\end{align}
where $P_\cap(i,j)$ is given by (\ref{moderna}).

%A rather tight interval can be determined where the ergodic spectral efficiency is sure to lie, precisely (see Appendix \ref{Topo})
%\be
%\E_\bH \big[  \cI(\SNR,\bH) \big] = \E_\bH \big[  \cH(\by) \big] - \E_\bH \big[  \cH(\by | \bx)  \big] ,
%\ee
%where
%\begin{align}
%\E_\bH \big[  \cH(\by | \bx)  \big] & = 2 \Nr \, \E_r \! \left[ \cH_{\rm b} \! \left( Q  \! \left( \! -   \sqrt{\SNR } \, r \! \right) \!  \right) \! \right] \\
%& = \frac{2 \Nr}{\sqrt{2 \pi}} \! \int_{-\infty}^\infty \! \cH_{\rm b} \! \left( Q  \! \left( \! -   \sqrt{\SNR } \, \xi \! \right) \!  \right) e^{-\xi^2/2} \, \mathrm{d}\xi
%\label{HospNens}
%\end{align}
%while
%\be
%\E_\bH \big[  \cH(\by) \big]  \leq 2 \Nr,
%\label{JJ}
%\ee
%satisfied with equality for $\SNR \to 0$, and
%\begin{align}
%& \E_\bH \big[  \cH(\by) \big]  \geq 2 \Nt - 2 \Nr \nonumber \\
%& \qquad\;   - \log_2 \! \Bigg[ \sum_{i=0}^\Nt
% \left( \!\!
%\begin{array}{c}
%\Nt  \\
%i 
%\end{array}
%\!\! \right)^{\!\! 2}
%\left( \frac{1}{4 \pi^2} \arccos^2 \! \left( \frac{2 i}{\Nt} -1 \right) \! \right)^{\! \Nr} \nonumber \\
%& \qquad\; + 2  \sum_{i=0}^{\Nt} \sum_{j=i+1}^{\Nt} 
%\left( \!\!
%\begin{array}{c}
%\Nt  \\
%i 
%\end{array}
%\!\! \right) \!
%\left( \!\!
%\begin{array}{c}
%\Nt  \\
%j 
%\end{array}
%\!\! \right)
% P^\Nr_\cap(i,j)
% \Bigg]   \label{XSM}
%\end{align}
%with $P_\cap(i,j)$ given by (\ref{moderna}).

%In contrast with the ergodic spectral efficiency itself, which becomes unwieldy even for modest $\Nt$ and $\Nr$,
The bounds specified by (\ref{Lluc1})--(\ref{moderna})
are readily computable even for very large numbers of antennas. For $\Nt=\Nr=64$, for instance, a direct evaluation of $\E_\bH \big[  \cI(\SNR,\bH) \big]$ would require the $ \cI(\SNR,\bH)$ for many realizations of $\bH$, with each such mutual information calculation involving over $10^{75}$ terms. In contrast, the bounds entail the single SNR-dependent integral in (\ref{Lluc1}) along with (\ref{XSM}), which does not depend on the SNR and can be precomputed; Table~\ref{LBHy} provides such precomputation for a range of antenna counts.
Also of interest is that the upper bound becomes exact for $\SNR \to 0$.

%\begin{figure*}
%\begin{align}
%\E_\bH \big[  \cH(\by) \big] &  \geq 2 \Nt - 2 \Nr   - \log_2 \! \Bigg[ \sum_{p=0}^\Nt
% \left( \!\!
%\begin{array}{c}
%\Nt  \\
%p 
%\end{array}
%\!\! \right)^{\!\! 2}
%\left( \frac{1}{4 \pi^2} \arccos^2 \! \left( \frac{2p}{\Nt} -1 \right) \! \right)^{\! \Nr} \!\!\! + 2 \,  \sum_{p=0}^{\Nt} \sum_{q=p+1}^{\Nt} 
%\!\! \left( \!\!
%\begin{array}{c}
%\Nt  \\
%p 
%\end{array}
%\!\! \right) \!
%\left( \!\!
%\begin{array}{c}
%\Nt  \\
%q 
%\end{array}
%\!\! \right)
% P^\Nr_\cap(p,q)
% \Bigg] 
% \label{XSM}
%\end{align}
%\end{figure*}

\begin{figure*}[b]
\begin{align}
\!\!  P_\cap (i,j) & = \frac{1}{2 \pi}  \int_0^\infty \!\!\!\! \int_0^\infty \text{erfc} \! \left( - \frac{ \gamma (\Nt-i-j) + \xi (i-j) }{\sqrt{i(\Nt-i)+j(\Nt-j)}}  \right) \text{erfc} \! \left(   \frac{  \gamma (i-j) - \xi (\Nt-i-j)  }{ \sqrt{i(\Nt-i)+j(\Nt-j)}}  \right)  e^{-2 \left( \gamma^2+\xi^2 \right)} \, \rmd\gamma \, \rmd\xi
\label{moderna}
\end{align}
\end{figure*}

\begin{table}
	\renewcommand{\arraystretch}{1.1}
	\caption{Lower bound on $\E_\bH \! \big[ \cH(\by) \big]$ as a function of $\Nt$ and $\Nr$.}
	\label{LBHy}
	\centering
	\begin{tabular}{ |l|c|c|c|c|c|c|c| } 
		\hline
		$\Nr \downarrow$ $\Nt \to$   & 1 & 2 & 4 & 8 & 16 & 32  \\
		\hline\hline
		1  & $2$ &   $2$      & $2$ & $2$ & $2$ & 2 \\ 
		\hline
		2  & $2$ & $3.02$ & $3.65$ & $3.85$ & $3.92$ & $3.96$ \\ 
		\hline
		4 & $2$ & $3.81$ & $6.07$ & $7.15$ & $7.57$ & $7.78$ \\
		\hline
		8 & $2$ & $4$  & $7.79$  & $12.35$ & $14.14$ & $15.03$ \\
		\hline
		16 & $2$ & $4$  & $8$  & $15.87$  & $24.81$ & $28.14$ \\
		\hline
		32 & $2$ & $4$  &  $8$ & $16$  & $31.96$  & $49.6$ \\
		\hline
	\end{tabular}
\end{table}

\begin{figure}[ht!]
	\centering
	\includegraphics[width=1\linewidth]{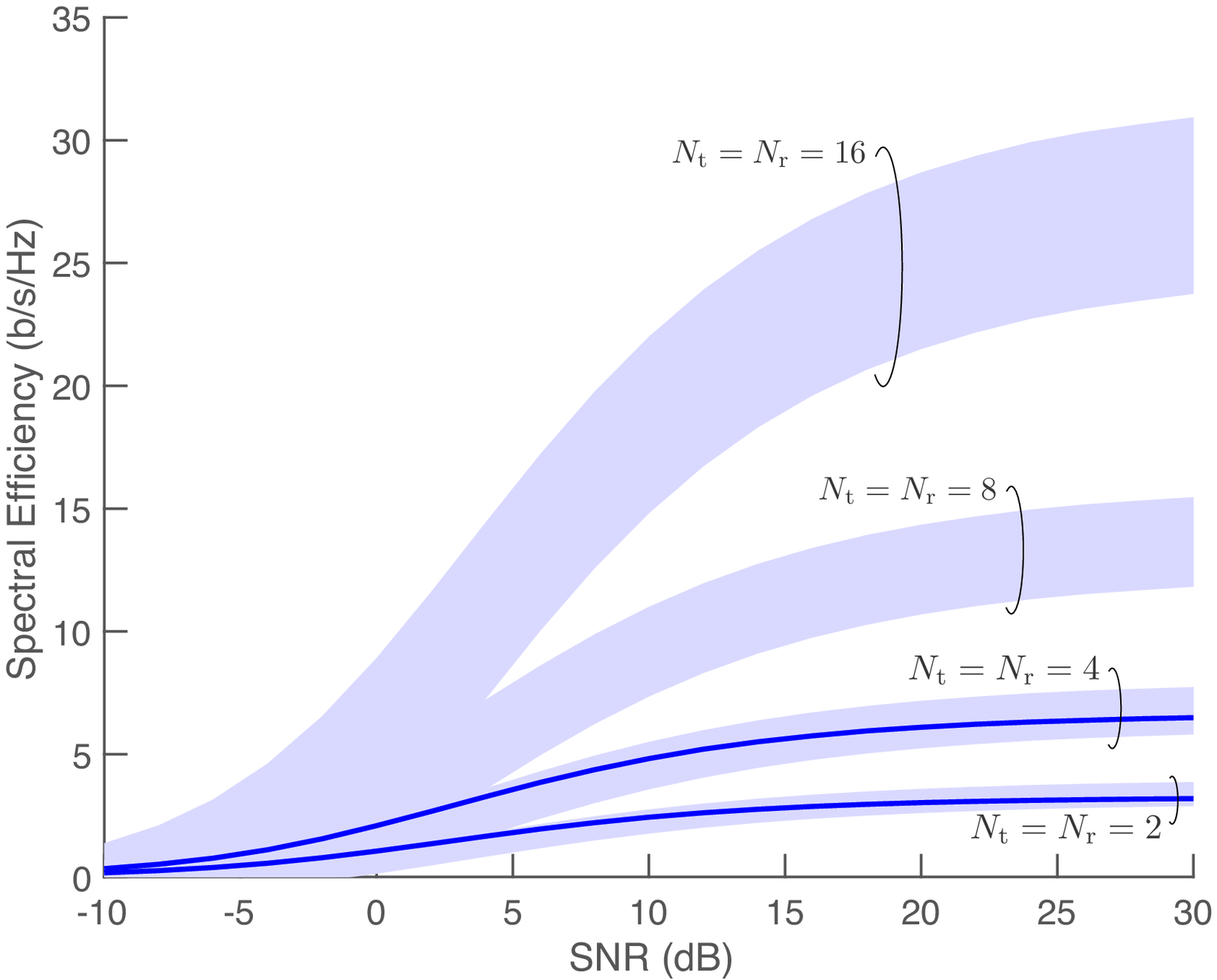}
	\caption{Ergodic spectral efficiency as a function of $\SNR$ with equiprobable signaling: the shaded areas are the bounding regions, the solid lines are the actual values for $\Nt=\Nr=2$ and $\Nt=\Nr=4$. The channel is IID Rayleigh-faded.}
	\label{C1thruC16approx}
\end{figure}

\begin{figure}[h]
	\centering
	\includegraphics[width=1\linewidth]{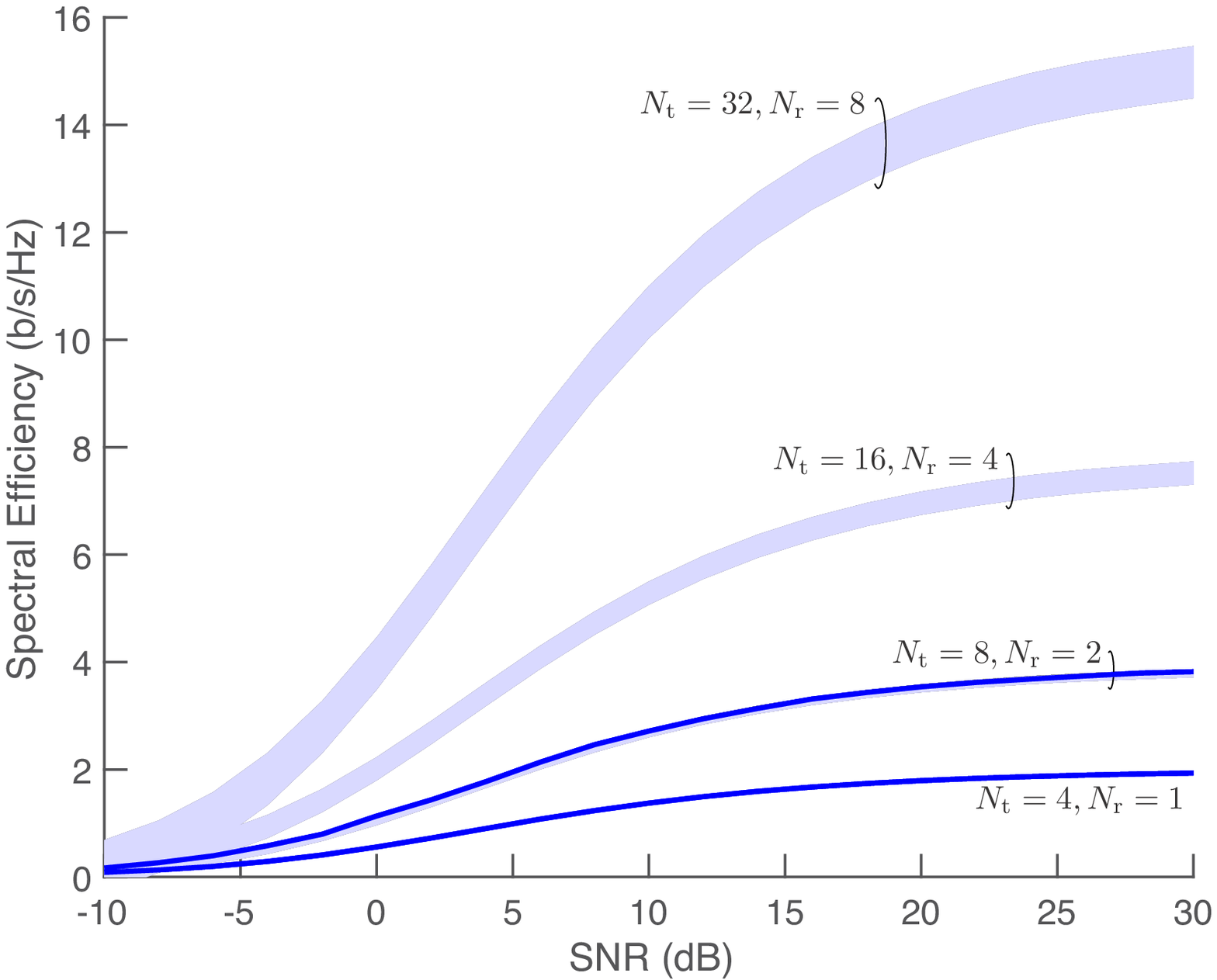}
	\caption{Ergodic spectral efficiency as a function of $\SNR$ with equiprobable signaling: the shaded areas are the bounding regions, the solid lines are the actual values for $\Nt=4,\Nr=1$ and $\Nt=8,\Nr=2$. The channel is IID Rayleigh-faded.}
	\label{Casymmetric}
\end{figure}

The range specified by the bounds is illustrated in Fig.~\ref{C1thruC16approx} for various values of $\Nt=\Nr$, alongside the actual spectral efficiencies (obtained via Monte-Carlo) for $\Nt=\Nr=2$ and $\Nt=\Nr=4$.

For $\Nt > \Nr$, the lower bound approaches its upper counterpart
%tightening the in-between range
and, for $\Nt \gg \Nr$, %it holds that
\begin{align}
& \E_\bH \big[  \cI(\SNR,\bH) \big] \label{JLA} \\
& \; \approx 2 \Nr \left( 1 -  \frac{1}{\sqrt{2 \pi}} \! \int_{-\infty}^\infty \! \cH_{\rm b} \! \left( Q  \! \left( \! -   \sqrt{\SNR } \, \xi \! \right) \!  \right) e^{-\xi^2/2} \, \mathrm{d}\xi  \right) \! .
\nonumber
\end{align}
Indeed, as detailed in Appendix \ref{Topo}, this approximation becomes an exact result for $\Nr=1$ or for $\Nt\to \infty$ with $\Nr$ arbitrary.

Some examples for $\Nt=4 \Nr$, presented in Fig.~\ref{Casymmetric}, confirm how precisely the ergodic spectral efficiency is determined when the antenna counts are somewhat skewed.

\section{Conclusion}
\label{calor7}

A host of issues that are thoroughly understood for full-resolution settings must be tackled anew for 1-bit MIMO communication.
In particular, the computation of the capacity becomes unwieldy for even very modest dimensionalities and the derivation of general precoding solutions becomes a formidable task, itself power-consuming. Fortunately, in the single-user case such general precoding can be circumvented via a judicious switching between beamforming and equiprobable signaling, with the added benefits that these transmissions strategies are much more amenable to analytical characterizations and that their requirements in terms of channel-state information at the transmitter are minimal: $\log_2 4^{\Nt-1}  =  4 \,(\Nt-1)$ bits for beamforming, none for equiprobable signaling.

The transition from beamforming to equiprobable signaling could be finessed by progressively activating quartets as the SNR grows, but the results in this paper suggest that there is a small margin of improvement: a direct switching at some appropriate point suffices to operate within a few dB of capacity at both low and intermediate SNRs.
It would be of interest to gauge this shortfall for more intricate channel models such as those in \cite{9405492,9411894,Undi2021}.

Channel estimation at the receiver is an important aspect, with the need for procedures that avoid having to painstakingly gauge all the transition probabilities between $\bx$ and $\by$ to deduce $\bH$. Of much interest would be to extend existing results for channel estimation with full-resolution DACs and 1-bit ADCs \cite{ivrlac2007mimo,7600443,li2017channel,wan2020generalized,atzeni2021channel}
to the complete 1-bit realm. Equally pertinent would be to establish the bandwidths over which a frequency-flat representation suffices for each channel model, and to extend the respective analyses to account for intersymbol interference. This is acutely important given the impossibility of implementing OFDM with 1-bit converters.

In those multiuser settings where orthogonal (time/frequency) multiple access is effective, switching between beamforming and equiprobable signaling is also enticing. In other cases, chiefly if the antenna numbers are highly asymmetric, orthogonal multiple access is decidedly suboptimum, and there is room for more general schemes. We hope that the results in this paper can serve as a stepping stone to such schemes.

% \cite{9007045,8754755,8487043,8811616,8331077,7967843,8103022,7472304,8462805,8010806}

%POINT TO CHANNEL ESTIMATION AND ADD REFS (\cite{7600443,li2017channel,wan2020generalized,atzeni2021channel}, MORE MISSING)
%NO CSIT NEEDED FOR EQUIPROBABLE, FEW BITS ONLY FOR BF. HENCE, THE LOW-SNR RESULTS FOR EQUIPROBABLE ARE OF INTEREST, EVEN IF SUBOPTIMUM, FOR WHENEVER THESE FEW BITS CANNOT BE CONVEYED.

%A MORE REFINED WAY OF TRANSITIONING FROM LOW-SNR BF TO HIGH-SNR EQUIPROBABLE COULD BE TO ACTIVATE AN INTERMEDIATE NUMBER OF QUARTETS, POSSIBLY IDENTIFIED USING THE PROCEDURE IN \cite{gao2018beamforming} (AND DENOTED THEREIN, WITH SOME ABUSE OF TERMINOLOGY, AS "BEAMFORMING" OVER A RANK-1 CHANNEL WITH HIGHER-ORDER CONSTELLATIONS). OR SOME BINARY-CONSTRAINED PRECODER COULD PERHAPS BE ENVISAGED. BUT MARGIN OF IMPROVEMENT SEEMS SMALL.

%FOR FUTURE WORK: EFFECTS OF PHASE NOISE, IQ IMBALANCE.
%SPATIAL-WIDENING \cite{8443598,8354789}.
%ALSO, TEST ON DETAILED CHANNEL  MODELS SUCH AS \cite{9405492,9411894,Undi2021}.

%FOR HOW TO COMPUTE LLRs AT THE RECEIVER, TO FEED SOFT-INPUT DECODERS, SEE \cite{gao2018beamforming}.

%\section*{Acknowledgment}
%
%This work was supported by the European Research Council under the H2020 Framework Programme/ERC grant agreement 694974,
%by Projects RTI2018-102112 and 101040, and by ICREA.

%-----------------------------------------------------------------------------------------------------

\appendices

\section{} %\section{Low-SNR Beamforming}
\label{superlliga}

Let $\sigma_0,\ldots,\sigma_{\Nt-1}$ be the singular values of $\bH$, ordered from largest to smallest, while $\bu_m$ and $\bv_m$ are the left and right singular vectors corresponding to $\sigma_m$.
From the singular value decomposition
\be
\bH = \!\!\! \sum_{m=0}^{\Nmin-1} \!\!\!  \sigma_m \bu_m \bv^*_m 
\ee
we have that
\be
\| \bH \bx \|^2 = \!\! \sum_{m=1}^{\Nmin-1} \!\! \sigma^2_m \left| \bv^*_m \bx \right|^2 ,
\label{dega}
\ee
which is the quantity to maximize. Under full-resolution transmission, (\ref{dega}) is maximized by $\bx \propto \bv_0$: complete projection on the dimension exhibiting the largest gain and zero projection elsewhere \cite[sec. 5.3]{Foundations:18}.
With 1-bit transmission, perfect alignment with $\bv_0$ is generally not possible, and the goal becomes to determine which $\bx$ best aligns. If $\bH$ is rank-1, then such $\bx$ is sure to maximize (\ref{dega}).
%If $\Nt \leq \Nr$, then again such $\bx$ maximizes (\ref{dega}) regardless of the channel rank, because any other $\bx$ will have an increased projection on lesser-gain dimensions at the expense of a reduced projection on the largest-gain dimension.
If the rank is plural, however, optimality cannot be guaranteed from best alignment with $\bv_0$ because some other $\bx$ leaning further away could have a more favorable projection across the rest of dimensions. Suppose, for instance, that the rank is $3$; if the $\bx$ best aligned with $\bv_0$ does not further project on $\bv_1$, but only on $\bv_2$, there could be another $\bx$ aligning slightly less with $\bv_0$ but projecting also on $\bv_1$ in a way that yields a higher metric in (\ref{dega}).
This possibility may arise when the largest singular value is not very dominant.
Even then, though, the $\bx$ that projects maximally on $\bv_0$ is bound to perform well.

Values of $\bx$ that align well with $\bv_0$ can be obtained as $\bx = \text{sgn}\big( e^{\j \varphi} \bv_0 \big)$ where $\varphi$ allows setting the absolute phase arbitrarily before quantization. Letting $\varphi  $ run from $0$ to $2 \pi$, every entry of the quantized $\bx$ changes four times and a subset of $4 \Nt$ vectors $\bx$ is obtained. These $4 \Nt$ vectors actually belong to $\Nt$ quartets because, if $\bsfx_k$ is in the $k$th subset, $\j \bsfx_k$ is sure to be there too. Since identifying one representative per quartet suffices for our purposes, attention can be restricted to those values of $\varphi$ that trigger a change in $\text{sgn}\big( e^{\j \varphi} \bv_0 \big)$, i.e., $\varphi = \angle(v_{0,m})$ for $m=0,\ldots,\Nt-1$. Letting $\varphi_m = \angle(v_{0,m}) + \epsilon$ with $\epsilon$ a small quantity, we obtain the subset of $\Nt$ quartet representatives as
\be
\bsfx_k =  \text{sgn}\big( e^{\j \varphi_{k-1}} \bv_0 \big) \qquad\quad k=1,\ldots,\Nt .
\label{picanteria}
\ee
The sign of $\epsilon$ is irrelevant, it merely changes which representative is selected for each quartet.
Confirming the intuition that the $\Nt$ quartets in (\ref{picanteria}) are good choices, it is proved in \cite{gao2018beamforming} that the quartet that best aligns with $\bv_0$  is sure to be in this subset. Thus, searching a subset of $\Nt$ candidates suffices to beamform optimally in rank-1 channels, and quasi-optimally in higher-rank channels, without having to search the entire field of $4^{\Nt-1}$ possibilities. 

Let us now turn to the performance.
%An upper bound on $\|  \bH \bsfx_{k^\star} \|^2$ can be obtained by maintaining the constraint of fixed magnitude on the entries of $\bx$, but relaxing the $90^\circ$ phase resolution. In particular, with arbitrary phase resolution, the optimum strategy would be to transmit $\bx = \sqrt{2} \, \bsfv_1$ where
%\be
%\bsfv_1 = \left[
%\begin{array}{ccc}
% e^{\j \phi_1} \vspace{-2mm} \\
%  \vdots \vspace{0.5mm} \\
% e^{\j \phi_{\Nt}}  
%\end{array}
%\right]
%\ee
%with $\phi_m = \angle (v_{1,m})$.
%Hence, with 1-bit transmission, using $\lambda_m = \sigma^2_m$,
%\be
%\|  \bH \bsfx_{k^\star} \|^2 \leq 2 \sum_{m=1}^{\Nt} \lambda_m \left| \bv_m \bsfv_1 \right|^2 ,
%\label{mar}
%\ee
%where, in contrast with a full-resolution transmission, some power may leak onto dimensions other than the one experiencing the strongest gain.
%Plugging (\ref{mar}) into (\ref{analitica}), the lower bound on $\ebnomininline$ is obtained.
An upper bound on $\|  \bH \bsfx_{k^\star} \|^2$ can be obtained by assuming that, on every channel realization, there is a value of $\bx$ that aligns perfectly with $\bv_0$. From (\ref{dega}), this gives
\be
\| \bH \bx \|^2 \leq 2 \Nt \sigma^2_0,
\label{consellers}
\ee
which, along with (\ref{Argimon2}), yields the lower bound in (\ref{bounds}).

In turn, a lower bound on $\|  \bH \bsfx_{k^\star} \|^2$ is obtained for any choice of $\bx$, and in particular for $\bx = \text{sgn} \big( e^{\j \varphi} \bv_0 \big)$ with $\varphi \in [-\pi/4,\pi/4]$, such that \cite{gao2018beamforming}
\begin{align}
\left| \bv^*_0 \bsfx_{k^\star} \right| & \geq \max_{\varphi \in [-\pi/4,\pi/4]} \left| \bv^*_0 \, \text{sgn} \big( e^{\j \varphi} \bv_0 \big) \right| \\
& \geq \E_\varphi \!\! \left[  \left| \bv^*_0 \, \text{sgn} \big( e^{\j \varphi} \bv_0 \big) \right| \right]  \\
& = \E_\varphi \!\! \left[  \left| \sum_{m=0}^{\Nt-1} v^*_{0,m} \, \text{sgn} \big( e^{\j \varphi} v_{0,m} \big) \right| \right] \\
& = \E_\varphi \!\! \left[  \left| \sum_{m=0}^{\Nt-1} \! \left|v_{0,m}\right| e^{-\j \phi_m} \, \text{sgn} \big( e^{\j (\varphi+\phi_m)} \big) \right| \right] \nonumber \\
& = \E_\varphi \!\! \left[  \left| \sum_{m=0}^{\Nt-1} \! \left|v_{0,m}\right| e^{\j \varphi} e^{-\j (\varphi+\phi_m)} \, \text{sgn} \big( e^{\j (\varphi+\phi_m)} \big) \right| \right] 
\nonumber \\
& = \E_\varphi \!\! \left[  \left| \sum_{m=0}^{\Nt-1} \! \left|v_{0,m}\right| e^{-\j (\varphi+\phi_m)} \, \text{sgn} \big( e^{\j (\varphi+\phi_m)} \big) \right| \right] \nonumber .
\end{align}
For any $\theta \in [0,2\pi]$, the phase of $e^{-\j \theta} \text{sgn} \big( e^{\j \theta} \big)$ is within $[-\pi/4,\pi/4]$ while $\left| e^{-\j \theta} \text{sgn} \big( e^{\j \theta} \big) \right| = \sqrt{2}$.
Hence, letting $\theta_m=\varphi+\phi_m$,
\begin{align}
\left| \bv^*_0 \bsfx_{k^\star} \right| &  \geq \sqrt{2} \, \E_{\theta_0,\ldots,\theta_{\Nt-1}} \!\! \left[  \left| \sum_{m=0}^{\Nt-1} \! \left|v_{0,m}\right| e^{-\j \theta_m} \right| \right] \\
& \geq \sqrt{2} \,  \, \E_{\theta_0,\ldots,\theta_{\Nt-1}} \!\! \left[  \left| \sum_{m=0}^{\Nt-1} \! \left|v_{0,m}\right| \cos(\theta_m) \right| \right] \\
& = \sqrt{2} \,   \sum_{m=0}^{\Nt-1} \! \left|v_{0,m}\right| \E \big[ \! \cos(\theta_m) \big] \label{CEC} \\
& = \sqrt{2} \,   \sum_{m=0}^{\Nt-1} \! \left|v_{0,m}\right| \frac{2}{\pi} \int_{-\pi/4}^{\pi/4} \!\! \cos(\xi) \, \mathrm{d}\xi \\
& = \frac{4}{\pi} \sum_{m=0}^{\Nt-1} \! \left|v_{0,m}\right| \\
& = \frac{4}{\pi} \, \| \bv_0 \|_1 ,
\end{align}
where (\ref{CEC}) holds because $\cos(\theta_m) > 0$ for $\theta_m \in [-\pi/4,\pi/4]$.
Disregarding $\left| \bv^*_m \bsfx_{k^\star} \right|$ for $m>0$ in (\ref{dega}),
\be
\|  \bH \bsfx_{k^\star} \|^2 \geq \frac{16}{\pi^2} \, \sigma^2_0 \,  \| \bv_0 \|^2_1 ,
\label{BarPlaca}
\ee
which, along with (\ref{Argimon2}), yields the upper bound in (\ref{bounds}).

%\section{}
%\label{Noah}
%
%Using
%$
%\bH \bx_{k^\star} \bx^*_{k^\star} \bH^* = \Nt \Nr \, |\bv^* \bx_{k^\star}|^2 \, \bu \bu^* 
%$
%in (\ref{consellers}) and (\ref{BarPlaca}), we have that
%\be
%\frac{256}{\pi^4} N^4_{\rm t} N^2_{\rm r} \leq \| \bH \bx_{k^\star} \|^4 \leq 4 N^4_{\rm t} N^2_{\rm r} 
%\label{Cholo1}
%\ee
%while
%\begin{align}
%\| \bH \bx_{k^\star} \|^4_4 & = N^2_{\rm t} N^2_{\rm r} \, \| \bu \|^4_4 \, |\bv \bx_{k^\star}|^4 \\
%& =  N^2_{\rm t} \Nr \, |\bv \bx_{k^\star}|^4  
%\end{align}
%satisfies
%\be
%\frac{256}{\pi^4} N^4_{\rm t} \Nr \leq \| \bH \bx_{k^\star} \|^4_4 \leq 4 N^4_{\rm t} \Nr .
%\label{Cholo2}
%\ee
%Also,
%\begin{align}
%\!\! \tr \big( ( \text{nondiag}(\bH \bx_{k^\star} \bx^*_{k^\star} \bH^*)   )^2  \big) & = N^2_{\rm t} \, |\bv^* \bx_{k^\star}|^4  \\
%& \quad \cdot \tr \big( ( \Nr \bu \bu^* - \bI )^2  \big) , \nonumber
%\end{align}
%satisfying
%\be
%\tr \big( ( \text{nondiag}(\bH \bx_{k^\star} \bx^*_{k^\star} \bH^*)   )^2  \big)  \leq 4 N^4_{\rm t} \Nr  (\Nr-1)
%\label{Cholo3}
%\ee
%and
%\be
%\tr \big( ( \text{nondiag}(\bH \bx_{k^\star} \bx^*_{k^\star} \bH^*)   )^2  \big)  \geq \frac{256}{\pi^4} N^4_{\rm t} \Nr  (\Nr-1) .
%\label{Cholo4}
%\ee
%Plugging (\ref{Cholo1}), (\ref{Cholo2}), (\ref{Cholo3}), and (\ref{Cholo4}) into (\ref{ForzaOriol}), the lower and upper bounds on $S_0$ are obtained.

\section{}
\label{Topo}

%Starting with
%\be
%\E_\bH \big[  \cI(\SNR,\bH) \big] = \E_\bH \big[  \cH(\by) \big] - \E_\bH \big[  \cH(\by | \bx)  \big] ,
%\ee
%we can
In (\ref{TrumpOut2}), for every $n$ and $k$, $\Re\{\bh_n \bsfx_k  \} \sim \cN(0,\Nt)$ and $\Im\{\bh_n \bsfx_k  \} \sim \cN(0,\Nt)$. Thus, letting $r \sim \cN(0,1)$, 
\begin{align}
\E_\bH \big[  \cH(\by | \bx)  \big] & = 2 \Nr \, \E_r \! \left[ \cH_{\rm b} \! \left( Q  \! \left( \! -   \sqrt{\SNR } \, r \! \right) \!  \right) \! \right] \\
& = \frac{2 \Nr}{\sqrt{2 \pi}} \! \int_{-\infty}^\infty \! \cH_{\rm b} \! \left( Q  \! \left( \! -   \sqrt{\SNR } \, \xi \! \right) \!  \right) e^{-\xi^2/2} \, \mathrm{d}\xi . \nonumber
\end{align}
In turn, 
\be
\E_\bH \big[  \cH(\by) \big]  \leq 2 \Nr
\label{Lluc3}
\ee
with equality for $\SNR \to 0$, when the receiver observes only noise and $\by$ is equiprobably binary on $2 \Nr$ real dimensions.
As the removal of noise can only decrease it, $\cH(\by)$ diminishes as the SNR grows, being lower-bounded by its value for $\SNR \to \infty$. The expectation of such noiseless lower bound over $\bH$ can be elaborated by generalizing to our complex setting a clever derivation in \cite{gao2017power}, starting from
\begin{align}
& \E_\bH \big[  \cH(\by) \big] = \E_\bH \!\! \left[ \sum_{\ell=1}^{4^{\Nr}} p_\by(\bsfy_\ell) \log_2 \frac{1}{p_\by(\bsfy_\ell)}  \right]  \\
 & \qquad = \E_\bH \!\! \left[ \sum_{\ell=1}^{4^{\Nr}} \E_\bx \big[ p_{\by|\bx}(\bsfy_\ell | \bx) \big] \log_2 \frac{1}{\E_\bx \big[ p_{\by|\bx}(\bsfy_\ell | \bx) \big]}  \right] \nonumber \\
 & \qquad = \E_\bH \!\! \left[ \sum_{\ell=1}^{4^{\Nr}} \E_\bx \big[ 1\{ \text{sgn}(\bH \bx) = \by_\ell \} \big] \right. \nonumber \\
& \qquad  \qquad\quad \cdot \log_2 \frac{1}{\E_\bx \big[ 1\{ \text{sgn}(\bH \bx) = \by_\ell \} \big]}  \Bigg]
\label{Aragones}
\end{align}
where $1\{ \cdot \}$ is the indicator function. Since $\bH$ %has IID circularly symmetric entries and it
is isotropic and $\bx$ is equiprobable, no $\bsfy_\ell$ is favored over the rest in terms of the probability of $\text{sgn}(\bH \bx)$ equalling such $\bsfy_\ell$. Hence, (\ref{Aragones}) can be evaluated for any specific 
$\bsfy_\ell$, say $\bsfy_1 $ whose entries all equal $1+\j$. This gives
\begin{align}
 \E_\bH \big[  \cH(\by) \big] & = 4^\Nr \,  \E_\bH \! \Bigg[  \E_\bx \big[ 1\{ \text{sgn}(\bH \bx) = \bsfy_1 \} \big]  \nonumber \\
& \quad  \cdot \log_2 \frac{1}{\E_\bx \big[ 1\{ \text{sgn}(\bH \bx) = \bsfy_1 \} \big]}  \Bigg] .
\end{align}
Likewise, the probability that $\text{sgn}(\bH \bx) = \bsfy_1$ is common to every realization of $\bx$ and thus
\begin{align}
\E_\bH \big[  \cH(\by) \big] & = - 4^\Nr \,  \E_\bH \! \Big[  1\{ \text{sgn}(\bH \bsfx_1) = \bsfy_1 \}   \nonumber \\
& \quad  \cdot \log_2 \E_\bx \big[ 1\{ \text{sgn}(\bH \bx) = \bsfy_1 \} \big]  \Big] 
\end{align}
where all entries of $\bsfx_1$ equal $1+\j$ and where it is convenient to retain the second expectation over $\bx$ in order to later solve its counterpart over $\bH$.
Then,
\begin{align}
\E_\bH \big[  \cH(\by) \big] & = - 4^\Nr \,  \E_{\bH | \text{sgn}(\bH \bsfx_1) = \bsfy_1 } \! \Big[  1\{ \text{sgn}(\bH \bsfx_1) = \bsfy_1 \}   \nonumber \\
& \quad  \cdot \log_2 \E_\bx \big[ 1\{ \text{sgn}(\bH \bx) = \bsfy_1 \} \big]  \Big] \nonumber \\
& \quad \cdot \mathbb{P} [\text{sgn}(\bH \bsfx_1) = \bsfy_1]
\end{align}
and, since $\mathbb{P} [\text{sgn}(\bH \bsfx_1) = \bsfy_1] = \frac{1}{4^\Nr}$ and the factor $1\{ \text{sgn}(\bH \bsfx_1) = \bsfy_1 \} $ becomes immaterial once the expectation over $\bH$ has been conditioned on $\text{sgn}(\bH \bsfx_1)$, 
\begin{align}
& \E_\bH \big[  \cH(\by) \big] \\
& \;  = -  \E_{\bH | \text{sgn}(\bH \bsfx_1) = \bsfy_1 } \! \Big[  \log_2 \E_\bx \big[ 1\{ \text{sgn}(\bH \bx) = \bsfy_1 \} \big]  \Big] \nonumber \\
 & \;  = -  \E_{\bH | \text{sgn}(\bH \bsfx_1) = \bsfy_1 } \!\! \left[  \log_2 \frac{1}{4^\Nt}  \sum_{k=1}^{4^\Nt}   1\{ \text{sgn}(\bH \bsfx_k) = \bsfy_1 \}  \! \right] \nonumber \\
 & \; = 2 \Nt  -  \E_{\bH | \text{sgn}(\bH \bsfx_1) = \bsfy_1 } \!\! \left[  \log_2  \sum_{k=1}^{4^\Nt}   1\{ \text{sgn}(\bH \bsfx_k) = \bsfy_1 \} \! \right] \nonumber \\
 & \; \geq 2 \Nt - \log_2 \sum_{k=1}^{4^\Nt} \E_{\bH | \text{sgn}(\bH \bsfx_1) = \bsfy_1 } \big[ 1\{ \text{sgn}(\bH \bsfx_k) = \bsfy_1 \} \big] \nonumber 
\end{align}
where the last step follows from Jensen's inequality. Since the expectation of an indicator function yields the probability of the underlying event,
\begin{align}
& \E_\bH \big[  \cH(\by) \big]  \geq 2 \Nt \label{astra}  \\
& \qquad\quad - \log_2 \sum_{k=1}^{4^\Nt} \bbP \big[ \text{sgn}(\bH \bsfx_k) = \bsfy_1    | \text{sgn}(\bH \bsfx_1) = \bsfy_1   \big]  \nonumber
\end{align}
with
\begin{align}
& \bbP \big[ \text{sgn}(\bH \bsfx_k) = \bsfy_1   | \text{sgn}(\bH \bsfx_1) = \bsfy_1   \big] \\
& \qquad\quad = \frac{\bbP \big[  \text{sgn}(\bH \bsfx_k) = \bsfy_1  \, \cap \, \text{sgn}(\bH \bsfx_1) = \bsfy_1   \big] }{\bbP \big[ \text{sgn}(\bH \bsfx_1) = \bsfy_1 \big]} \nonumber \\
& \qquad\quad = 4^\Nr \,  \bbP \big[  \text{sgn}(\bH \bsfx_k) = \bsfy_1  \, \cap \, \text{sgn}(\bH \bsfx_1) = \bsfy_1   \big]  . \nonumber
\end{align}

As the channel has IID entries, letting $\bh$ be an arbitrary row of $\bH$,
\be
\bbP \big[  \text{sgn}(\bH \bsfx_k) = \bsfy_1  \, \cap \, \text{sgn}(\bH \bsfx_1) = \bsfy_1   \big]  = (P_\cap)^\Nr 
\ee
with
\begin{align}
P_\cap & = \bbP \big[  \text{sgn}(\bh \bsfx_k) = (1+\j) \, \cap \, \text{sgn}(\bh \bsfx_1) = (1+\j)   \big] \nonumber \\
& = \bbP \big[  \text{sgn}( \Re\{ \bh \bsfx_k \}) = 1 \, \cap \, \text{sgn}( \Im\{ \bh \bsfx_k \}) = 1 \nonumber \\
& \quad  \, \cap \,  \text{sgn}( \Re\{ \bh \bsfx_1 \}) = 1 \, \cap \, \text{sgn}( \Im\{ \bh \bsfx_1 \}) = 1     \big]  \\
& = \bbP \big[   \Re\{ \bh \bsfx_k \} >0 \, \cap \,  \Im\{ \bh \bsfx_k \} >0 \nonumber \\
& \quad  \, \cap \,   \Re\{ \bh \bsfx_1 \} >0 \, \cap \,  \Im\{ \bh \bsfx_1 \} >0     \big] \label{Asens} . 
\end{align}
Defining $a_k=\bh \bsfx_k $ and $a_1=\bh \bsfx_1$,
\begin{align}
P_\cap & = \int_0^\infty \!\!\!\! \int_0^\infty \!\!\!\! \int_0^\infty \!\!\!\! \int_0^\infty f_{\Re\{a_k\}\Im\{a_k\}\Re\{a_1\}\Im\{a_1\}   } (\alpha,\beta,\gamma,\xi) \nonumber \\
& \quad\; \cdot \rmd\alpha \, \rmd\beta \,\rmd\gamma \, \rmd\xi
\label{mosquiteres}
\end{align}
with $\Re\{a_k\}$, $\Im\{a_k\}$, $\Re\{a_1\}$, and $\Im\{a_1\} $ jointly Gaussian with mean zero and covariance
\be
\bSigma_k = 
\left[
\begin{array}{cc}
\Nt \bI  & \bR_k  \\
\bR_k  & \Nt \bI 
\end{array}
\right]
\ee
where $\bI$ is the $2 \times 2$ identity matrix while
\be
\bR_k = 
\left[
\begin{array}{cc}
\Nt - i -j  & j-i   \\
i-j  & \Nt-i-j 
\end{array}
\right] .
\ee
with $i$ and $j$ the respective number of entries of $\Re\{a_k\}$ and $\Im\{a_k\}$ that are $-1$, the rest of their entries (along with all the entries of $\Re\{a_1\}$ and $\Im\{a_1\}$) being $+1$. Most importantly, because the entries of $\bh$ are IID, the position of those $-1$ values is immaterial and only their totals $i$ and $j$ matter. Altogether, the joint distribution of $\Re\{a_k\}$, $\Im\{a_k\}$, $\Re\{a_1\}$ and $\Im\{a_1\} $ is as in (\ref{pfizer}) and, plugging it in (\ref{mosquiteres}) and tediously
integrating over two of the dimensions, what emerges is (\ref{moderna}) with the dependence on $i$ and $j$ made explicit and with $\text{erfc}(\cdot)$ the complementary error function.

Returning to (\ref{astra}), and accounting for the number of indices $k$ that map to each $i$ and $j$,
\begin{align}
 \E_\bH \big[  \cH(\by) \big] & \geq 2 \Nt - 2 \Nr \label{flors} \\
& \quad - \log_2 \sum_{i=0}^{\Nt} \sum_{j=0}^{\Nt} 
\left( \!\!
\begin{array}{c}
\Nt  \\
i 
\end{array}
\!\! \right) \!
\left( \!\!
\begin{array}{c}
\Nt  \\
j 
\end{array}
\!\! \right)
P^\Nr_\cap(i,j) .  \nonumber
\end{align}
%For $p=q=0$ and $p=q=\Nt$, the covariance $\bSigma$ is singular and (\ref{moderna}) does not apply, but it is straightforward to see that $P_\cap(0,0) = 1/4$ and $P_\cap(\Nt,\Nt)=0$.
For $i=j$, (\ref{moderna}) can be integrated into
\be
P_\cap(i,i)= \frac{1}{4 \pi^2} \arccos \! \left( \frac{2 i }{\Nt} -1 \right)
\ee
with $P_\cap(0,0)=1/4$ and $P_\cap(\Nt,\Nt)=0$.
Furthermore, $P_\cap(i,j)=P_\cap(j,i)$ with $P_\cap(0,\Nt)=P_\cap(\Nt,0)=0$.
With these relationships accounted for, (\ref{flors}) yields (\ref{XSM}).

%such that, altogether,
%\begin{align}
%& \E_\bH \big[  \cH(\by) \big]  \geq 2 \Nt - 2 \Nr \nonumber \\
%& \qquad\;   - \log_2 \! \Bigg[ \sum_{p=0}^\Nt
% \left( \!\!
%\begin{array}{c}
%\Nt  \\
%p 
%\end{array}
%\!\! \right)^{\!\! 2}
%\left( \frac{1}{4 \pi^2} \arccos^2 \! \left( \frac{2p}{\Nt} -1 \right) \! \right)^{\! \Nr} \nonumber \\
%& \qquad\; + 2  \sum_{p=0}^{\Nt} \sum_{q=p+1}^{\Nt} 
%\left( \!\!
%\begin{array}{c}
%\Nt  \\
%p 
%\end{array}
%\!\! \right) \!
%\left( \!\!
%\begin{array}{c}
%\Nt  \\
%q 
%\end{array}
%\!\! \right)
% P^\Nr_\cap(p,q)
% \Bigg] .  \nonumber
%\end{align}

\begin{figure*}
\begin{align}
f_{\Re\{a_k\}\Im\{a_k\}\Re\{a_1\}\Im\{a_1\}   } (\alpha,\beta,\gamma,\xi) = \frac{1}{8 \pi^2} \frac{1}{i(\Nt-i)+j(\Nt-j)}
\exp \! \left(-\frac{1}{4} \frac{[\alpha \; \beta \; \gamma \; \xi] \, \bSigma^{-1}_k \, [\alpha \; \beta \; \gamma \; \xi]^{\rm T}}{i(\Nt-i)+j(\Nt-j)}  \right)
\label{pfizer}
\end{align}
\end{figure*}

Let us now consider some special cases of interest. For $\Nt=1$, (\ref{XSM}) reduces to
\be
\E_\bH \big[ \cH(\by) \big] \geq 2 - 2\Nr - \log_2 \! \left( 4^{-\Nr} + 4 P_\cap(0,1) \right) 
\ee
and, since $P_\cap(0,1)=0$, further to $\E_\bH \big[ \cH(\by) \big] \geq 2$.
In fact, in this case $\cH(\by) = 2$ for every channel realization and thus $\E_\bH \big[ \cH(\by) \big] = 2$.
%Combined with (\ref{Lluc3}), this gives $\E_\bH \big[ \cH(\by) \big] = 2$.

For $\Nr=1$, the scalar quantized signal $y$ takes four equiprobable values---again, not only on average, but for every channel realization---and thus $\E_\bH \big[ \cH(y) \big] = 2$.

For fixed $\Nt$ and $\Nr \to \infty$, the key is the observation that $P_\cap(i,j)$ achieves its largest value for $i=j=0$, namely $P_\cap(0,0)=1/4$. For $i>0$ and/or $j>0$, $P_\cap(i,j) < 1/4$ because any negative sign in either the real and imaginary parts of $\bsfx_k$ reduces the probability in (\ref{Asens}). 
%The smallest value is, as advanced, $P_\cap(\Nt,\Nt)=0$. 
The largest term in the summations within the logarithm in (\ref{XSM}) equals $4^{-\Nr}$ and, as $\Nr \to \infty$, every other term vanishes faster and the lower bound on $\E_\bH \big[ \cH(\by) \big]$ converges towards $2 \Nt$.

Finally, for fixed $\Nr$ and $\Nt \to \infty$, the rows of $\bH$ become asymptotically orthogonal \cite[sec. 5.4.2]{Foundations:18} and hence, for every realization of $\bH$, $\by$ consists of IID complex components. Again, $\E_\bH \big[ \cH(\by) \big] = 2 \Nr$.

For $\Nr = 1$ and for $\Nt \to \infty$ with fixed $\Nr$, the above observations reveal that the lower and upper bounds coincide, fully determining, as per (\ref{JLA}), the ergodic spectral efficiency with equiprobable signaling and IID Rayleigh fading.

\bibliographystyle{IEEEtran}
\bibliography{jour_short,conf_short,library2,thz_refs,info_theory_refs}

\end{document}